\documentclass[preprint,prd,showpacs,preprintnumbers,amsmath,amssymb,eqsecnum,nofootinbib]{revtex4}

\usepackage{graphicx}% Include figure files
\usepackage{dcolumn}% Align table columns on decimal point
\usepackage{bm}% bold math
\newcommand{\bea}{\begin{eqnarray}}
\newcommand{\eea}{\end{eqnarray}}
\newcommand{\bi}{\begin{itemize}}
\newcommand{\ei}{\end{itemize}}
\newcommand{\benu}{\begin{enumerate}}
\newcommand{\eenu}{\end{enumerate}}
\newcommand{\nn}{\nonumber}
\begin{document}

\preprint{KEK-CP-215\quad
YNU-HEPTh-09-102\quad
KUNS-2189}

\title{Parton distributions in the virtual photon target\\
up to NNLO in QCD}% Force line breaks with \\

\author{Takahiro UEDA}
\email{uedat@post.kek.jp}
\affiliation{ High Energy Accelerator Research Organization (KEK),
  1-1 Oho, Tsukuba, Ibaraki 305-0801, Japan
}%

\author{Ken SASAKI}
 \email{sasaki@phys.ynu.ac.jp}
\affiliation{
Dept. of Physics,
Faculty of Engineering, Yokohama National University,
 Yokohama 240-8501, Japan
}%

\author{Tsuneo UEMATSU}
\email{uematsu@scphys.kyoto-u.ac.jp}
\affiliation{Dept. of Physics, Graduate School of Science, Kyoto University,
 Yoshida, Kyoto 606-8501, Japan
\vspace{1cm}}%


\begin{abstract}
Parton distributions in the virtual
photon target are investigated in perturbative QCD up to the next-to-next-to-leading order (NNLO).
In the case $\Lambda^2 \ll P^2 \ll Q^2$, where $-Q^2$ ($-P^2$) is the mass
squared of the probe (target) photon, parton distributions can be predicted
completely up to the NNLO, but they are factorisation-scheme-dependent.
We analyse parton distributions in two different factorisation schemes,
namely $\overline{\rm MS}$ and ${\rm DIS}_{\gamma}$ schemes,
and discuss their scheme dependence.
We show that the factorisation-scheme dependence is characterised
by the large-$x$ behaviours of quark distributions.
Gluon distribution is predicted to be very small in absolute value except in
the small-$x$ region.
\end{abstract}

\maketitle

%%%%%%%%%%%%%%%%%%%%%%%%%%%
\section{Introduction}
\begin{figure}
\begin{center}
\includegraphics[scale=0.4]{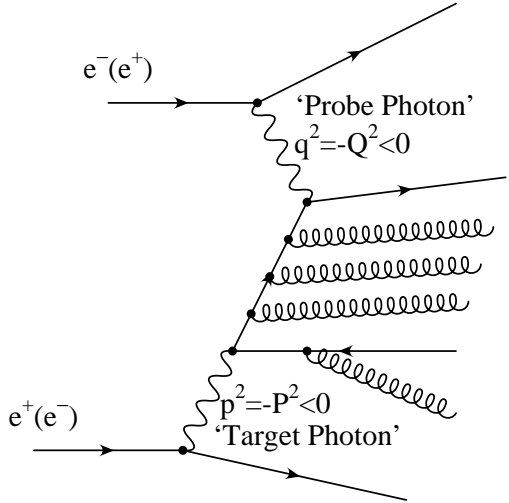}
\vspace{-0.5cm}
\caption{\label{Two Photon Process} Deep inelastic scattering on a
virtual photon in the $e^+~e^-$
collider experiments.}
\end{center}
\end{figure}

%%%%%%%%%%%%%%%%%%%%%%%%%%%%%
It is well known that, in $e^+ e^-$ collision experiments, the cross section
for the two-photon processes $e^+ e^- \rightarrow e^+ e^- + {\rm hadrons}$
illustrated in Fig.\ref{Two Photon Process} dominates at high energies over other processes such as the
one-photon annihilation process $e^+ e^- \rightarrow \gamma^*
\rightarrow {\rm hadrons}$. The two-photon processes are observed in the
double-tag events where both the outgoing $e^+$ and $e^-$ are detected.
Especially, the case in which one of the virtual photon
is far off-shell (large $Q^2\equiv -q^2$), while the other is close to
the mass-shell (small $P^2=-p^2$), can be viewed as a
deep-inelastic scattering where the target is a photon rather than a nucleon~\cite{WalshBKT}.
In this deep-inelastic scattering off photon targets, we can study the photon structure
functions, which are the analogues of the nucleon structure functions.

%%%%%%%%%%%%%%%%%%%%%%%%%%%%%%%%%%%%%
%%%%%%%%%%%%%%%%%%%%%%%%%%%%%%%%%%

A unique and interesting feature of the photon structure functions is
that, in contrast with the nucleon case, the target mass squared $P^2$
is not fixed but can take various values and that the structure functions show
different behaviours depending on the values of $P^2$.

In the case of a real photon target ($P^2=0$), unpolarised (spin-averaged) photon
structure functions $F_2^\gamma(x,Q^2)$ and $F_L^\gamma(x,Q^2)$ were studied first in the
parton model \cite{WalshZerwas}, and then investigated in perturbative QCD (pQCD).
In the framework based on the operator product expansion (OPE)
\cite{CHM} supplemented by the renormalisation (RG) group method,
Witten \cite{Witten} obtained the leading order (LO) QCD contributions to
$F_2^\gamma$ and $F_L^\gamma$ and, shortly after,
the next-to-leading order (NLO) QCD contributions were calculated by Bardeen
and Buras \cite{BB}.
The same results were rederived by the QCD improved parton model powered by
the parton evolution equations \cite{Dewitt,GR}.
Recently, some of the parameters which are necessary to evaluate
the next-to-next-to-leading order (NNLO) corrections to $F_2^\gamma$ were calculated
and also the NNLO solution of the evolution equations for the real photon case was dealt with~\cite{MVV}.
When polarised beams are used in $e^+e^-$ collision experiments, we can get information on the
spin structure of the photon.
The QCD analysis of the polarised photon structure function
$g_1^\gamma(x,Q^2)$ for the real photon target was performed in the LO \cite{KS} and in the NLO
\cite{SV,GRS}. For more information on the recent theoretical and experimental
investigation of unpolarised and polarised photon structure, see the review
articles \cite{Krawczyk}.

For a virtual photon target ($P^2\ne 0$), we obtain the
virtual photon structure functions $F_2^\gamma(x,Q^2,P^2)$ and $F_L^\gamma(x,Q^2,P^2)$.
They were studied in the LO~\cite{UW1} and in the
NLO~\cite{UW2} by pQCD. In fact, these structure functions were analysed
in the kinematical region,
\begin{equation}
\Lambda^2 \ll P^2 \ll Q^2~, \label{Kinematical region}
\end{equation}
where $\Lambda$ is the QCD scale parameter. The advantage of studying a virtual
photon target in this kinematical region (\ref{Kinematical region}) is that
we can calculate the whole structure function, its shape and magnitude,
by the perturbative method. This is contrasted with the case of the real photon target
where in the NLO and beyond there appear nonperturbative pieces.
The virtual photon structure functions $F_2^\gamma$ and $F_L^\gamma$
were also studied by the QCD improved parton model~\cite{Rossi,DG,GRStratmann,Fontannaz}.
In the same kinematical region (\ref{Kinematical region}),
the polarised virtual photon structure function $g_1^\gamma(x,Q^2,P^2)$ was
investigated up to the NLO in pQCD~\cite{SU1}, and its target mass effects
were studied~\cite{BSU}.
%%%%%%%%%%%%%%%%%%%%%%%%%%%
% and in the second paper of~\cite{GRS}.
%%%%%%%%%%%%%%%%%%%%%%%%%%%
Moreover, the first moment of $g_1^\gamma(x,Q^2,P^2)$
was calculated up to the NNLO~\cite{SUU}.

Recently the three-loop calculations were made for the anomalous dimensions of
both quark and gluon operators~\cite{MVV1,MVV2} and also for the photon-quark
and photon-gluon splitting functions~\cite{MVV3}.
Using these results, we analysed the virtual photon structure functions
$F_2^\gamma(x,Q^2,P^2)$ and $F_L^\gamma(x,Q^2,P^2)$ in the kinematical region
(\ref{Kinematical region}) in QCD. We could give
definite predictions for $F_2^\gamma(x,Q^2,P^2)$ up to the NNLO and for
$F_L^\gamma(x,Q^2,P^2)$ up to the NLO~\cite{USU}, and also for the target mass effects
on these functions~\cite{KSUU}.

In parton picture, structure functions are expressed as
convolutions of coefficient functions and parton distributions in the target.
And these parton distributions will be used for predicting the behaviours of
other processes. When the target is a virtual photon with $P^2$ being
in the kinematical region (\ref{Kinematical region}),
then a definite prediction can be made for its parton distributions in pQCD.
However, it is well known that parton distributions are dependent on the scheme
which is employed to factorise structure functions into
coefficient functions and parton distributions.
It is possible that parton distributions obtained
in one factorisation scheme may be more appropriate to use than those acquired in other schemes.
Indeed, for the case of the {\it polarised} virtual photon target,
the NLO QCD analysis was made for its parton distributions in
several factorisation schemes and their scheme dependence was closely studied~\cite{SU2}.

In this paper we perform the QCD analysis of the parton distributions in the
{\it unpolarised} virtual photon target up to the NNLO\footnote{This work is a sister version of Ref.\cite{SU2}.}. We investigate
the distributions of the flavour singlet and nonsinglet quarks and gluons
in two factorisation schemes, namely $\overline {\rm MS}$ and
${\rm DIS}_{\gamma}$ schemes\footnote{Part of the results in this paper was reported
 at the 8th International Symposium on Radiative Corrections (RADCOR)\cite{UUS}.}.
Using the framework of the QCD improved parton model, we give,
in the next section, the explicit expressions for the moments of
the flavour singlet (nonsinglet) quark and gluon distributions
up to the NNLO. In Sec.\ref{RenormSchemeDependence}, we show first that the parton distributions
obtained up to the NNLO are independent of the renormalisation-prescription
which is chosen in defining
the QCD coupling constant $\alpha_s$.
And then we explain the two factorisation schemes, $\overline {\rm MS}$ and
${\rm DIS}_{\gamma}$ schemes, which we consider in this paper.
The behaviours of the parton distributions near $x=1$ and their factorisation-scheme dependence
are discussed in Sec.\ref{LargeX}.
The numerical analysis of the parton distributions predicted by $\overline {\rm MS}$ and
${\rm DIS}_{\gamma}$ schemes will be given in Sec.\ref{NumericalAnalysis}.
The final section is devoted to the conclusion.

%%%%%%%%%%%%%%%%%%%%%%%
\section{Parton distributions in the virtual photon
\label{PartonDistribution}}
%%%%%%%%%%%%%%%%%%%%%%%

In this section
the parton distributions in the virtual photon are analysed in the framework of the QCD improved parton model~\cite{Altarelli} with the DGLAP parton evolution equations.

Let $q^i_{\pm}(x,Q^2,P^2)$, $G^{\gamma}_{\pm}(x,Q^2,P^2)$,
$\Gamma^{\gamma}_{\pm}(x,Q^2,P^2) $ be quark with $i$-flavour, gluon, and photon
distribution functions with $\pm$ helicities in the virtual
photon with mass $-P^2$. Then the spin-averaged parton distributions
in the virtual photon are defined as
\bea
   q^i&\equiv& \frac{1}{2} [q^i_+ +{\overline q}^i_+ +q^i_- +{\overline q}^i_-] ~,
\nonumber \\
G^{\gamma}&\equiv& \frac{1}{2}[G^{\gamma}_+ +G^{\gamma}_-]~, \qquad
\Gamma^{\gamma}\equiv \frac{1}{2}[\Gamma^{\gamma}_+ +\Gamma^{\gamma}_-] ~.
\eea
In the leading order of the electromagnetic coupling constant,
$\alpha=e^2/4\pi$, $\Gamma^{\gamma}$ does not evolve with $Q^2$ and is set to be
\begin{equation}
\Gamma^{\gamma}(x,Q^2,P^2)=\delta(1-x)~. \label{PhotonDistribution}
\end{equation}
%%%%%%%%%%%%%%%%%%%
Instead of using the quark distributions $q^i$, it is advantageous to treat the flavour singlet and nonsinglet combinations defined as follows:
\bea
    q^{\gamma}_S &\equiv& \sum_i~ q^i ~, \\
   q^{\gamma}_{NS} &\equiv& \sum_i  e^2_i \Bigl( q^i - \frac{q^{\gamma}_S}{n_f}
\Bigr)~,  \label{qNS}
\eea
where $n_f$ is the number of relevant active quark (i.e., the massless quark)
flavours and $e_i$ is the electromagnetic charge of the active quark with flavour $i$ in the unit of proton charge.

Now introducing a row vector
\begin{equation}
\bm{q}^\gamma=( q^\gamma_S,~G^\gamma,~q^\gamma_{NS}),\label{PartonRowVector}
\end{equation}
the parton distributions $\bm{q}^\gamma(x,Q^2,P^2)$ in the virtual photon
satisfy an inhomogeneous evolution equation \cite{Dewitt,GR,MVV}

\begin{equation}
\frac{d~\bm{q}^\gamma(x,Q^2,P^2)}{d~ \ln Q^2}=
\bm{k}(x,Q^2) +
\int_x^1\frac{dy}{y}\bm{q}^\gamma(y,Q^2,P^2) P(\frac{x}{y},Q^2)~,
\label{evol}
\end{equation}
where the elements of a row vector $\bm{k}= ( k_S,~
k_G,~  k_{NS})$ refer to the splitting functions of $\gamma$ to the singlet quark
combination,  to gluon and
 to nonsinglet quark combination, respectively.
The $3\times 3$ matrix $ P(z,Q^2)$ is expressed as
\bea
 P(z,Q^2)=\begin{pmatrix}
 P^S_{q q}(z,Q^2)& P_{G q}(z,Q^2)&0\\
            P_{q G}(z,Q^2)& P_{G G}(z,Q^2)&0 \\
    0&0& P^{NS}_{q q}(z,Q^2)\end{pmatrix}~, \label{SplitPMat}
\eea
where $P_{AB}$ is a splitting function of $B$-parton to $A$-parton.

There are two methods to obtain the parton distribution
$\bm{q}^\gamma(x,Q^2,P^2)$ up to the NNLO.
In one method, we use the parton splitting functions up to the three-loop level and
we solve numerically $\bm{q}^\gamma(x,Q^2,P^2)$ in Eq.(\ref{evol})
by iteration, starting
from the initial quark and gluon distributions of the virtual photon at
$Q^2=P^2$. The interesting point of studying the virtual photon with mass
$-P^2$ is that when $P^2 \gg \Lambda ^2$, the initial parton distributions
of the photon are completely known up to the two-loop level in QCD.
The other method, which is more common than the former, is by making use of
the inverse Mellin transformation. From now on we follow the latter method.
First we take the Mellin moments of Eq.(\ref{evol}),
\begin{equation}
\frac{d~\bm{q}^\gamma(n,Q^2,P^2)}{d~ \ln Q^2}=
\bm{k}(n,Q^2) +
\bm{q}^\gamma(n,Q^2,P^2) P(n,Q^2)~,
\label{Momentevol}
\end{equation}
where we have defined the moments of an arbitrary function $f(x)$ as
\begin{equation}
     f(n) \equiv \int_0^1 dx x^{n-1} f(x).
\end{equation}

Henceforce we omit the obvious
$n$-dependence for simplicity.
Expansions are made for the splitting functions $\bm{k}(Q^2)$ and $ P(Q^2)$
in powers of the QCD and QED coupling constants as follows:
\bea
\bm{k}(Q^2)&=&\frac{\alpha}{2\pi}\bm{k}^{(0)}+
\frac{\alpha\alpha_s(Q^2)}{(2\pi)^2}\bm{k}^{(1)} +
\frac{\alpha}{2\pi}\left[\frac{\alpha_s(Q^2)}{2\pi}\right]^2\bm{k}^{(2)}\cdots~, \label{k-expansion}\\
 P(Q^2)&=&\frac{\alpha_s(Q^2)}{2\pi} P^{(0)}
+\left[\frac{\alpha_s(Q^2)}{2\pi}\right]^2  P^{(1)}
+\left[\frac{\alpha_s(Q^2)}{2\pi}\right]^3  P^{(2)}
+\cdots  ~,  \label{P-expansion}
\eea
and a new variable $t$ is introduced as the evolution variable
instead of $Q^2$~\cite{FP1},
\begin{equation}
t \equiv \frac{2}{\beta_0}\ln\frac{\alpha_s(P^2)}{\alpha_s(Q^2)}~.
\end{equation}

The solution $\bm{q}^\gamma(t)(=\bm{q}^\gamma(n,Q^2,P^2))$ of Eq.(\ref{Momentevol}) is decomposed in the following form:
\begin{equation}
\bm{q}^\gamma(t)=\bm{q}^{\gamma(0)}(t)+\bm{q}^{\gamma(1)}(t)+\bm{q}^{\gamma(2)}(t)~,
\end{equation}
where the first, second and third terms represent the solution in the LO, NLO
and NNLO, respectively.
%%%%%%%%%%%%%%%%%%%%%%%%%%%%%
Then they satisfy the following three differential equations:
\bea
 \frac{d {\bm q}^{\gamma(0)}(t)}{d t}&=& \frac{\alpha}{\alpha_s(t)}{\bm k}^{(0)} +
     {\bm q}^{\gamma(0)}(t) P^{(0)}~,
\label{APELeading}\\
 \frac{d {\bm q}^{\gamma(1)}(t)}{d t}&=& \frac{\alpha}{2\pi}~\bm{R}_{K(1)} +
  \frac{\alpha_s(t)}{2\pi}~{\bm q}^{\gamma(0)}(t)~R_{P(1)} +
     {\bm q}^{\gamma(1)}(t) P^{(0)}~,
\label{APENext} \\
\frac{d {\bm q}^{\gamma(2)}(t)}{d t}&=& \frac{\alpha}{2\pi}
\frac{\alpha_s(t)}{2\pi}~\bm{R}_{K(2)}
+\Bigl(\frac{\alpha_s(t)}{2\pi}\Bigr)^2~{\bm q}^{\gamma(0)}(t)~R_{P(2)} \nonumber \\
&&\qquad +\frac{\alpha_s(t)}{2\pi}~{\bm q}^{\gamma(1)}(t)~R_{P(1)} +
     {\bm q}^{\gamma(2)}(t) P^{(0)}~,
\label{APENextNext}
\eea
where
\bea
\bm{R}_{K(1)}&=& \bm{k}^{(1)}-\frac{\beta_1}{2\beta_0} \bm{k}^{(0)}~, \label{Rv_K1} \\
   R_{P(1)}&=& P^{(1)}-\frac{\beta_1}{2\beta_0} P^{(0)}~, \label{R_P1} \\
\bm{R}_{K(2)}&=& {\bm{k}}^{(2)} - \frac{\beta_1}{2\beta_0}{\bm{k}}^{(1)}
+\frac{1}{4}\Bigl((\frac{\beta_1}{\beta_0})^2- \frac{\beta_2}{\beta_0}  \Bigr)
{\bm{k}}^{(0)}~, \label{Rv_K2} \\
   R_{P(2)}&=&  P^{(2)} - \frac{\beta_1}{2\beta_0} P^{(1)} +
\frac{1}{4}\Bigl((\frac{\beta_1}{\beta_0})^2- \frac{\beta_2}{\beta_0}  \Bigr)
 P^{(0)}~. \label{R_P2}
\eea
%%%%%%%%%%%%%%%%%%%%%%%%%%%%%%
where we have used the fact the QCD effective coupling constant
$\alpha_s(Q^2)$ satisfies
\begin{equation}
  \frac{d \alpha_s(Q^2)}{d {\rm ln} Q^2}= -\beta_0 \frac{\alpha_s(Q^2)^2}{4\pi}
   -\beta_1 \frac{\alpha_s(Q^2)^3}{(4\pi)^2}-
\beta_2 \frac{\alpha_s(Q^2)^4}{(4\pi)^3}+
\cdots ,
\label{beta}
\end{equation}
where $\beta_0=11-\frac{2}{3}n_f$~ and $\beta_1$ and $\beta_2$ were known
~\cite{JC,TVZ}.
Note that the $P^2$ dependence of $\bm{q}^\gamma$ solely comes from the
initial condition (or boundary condition) as we will see below.

The initial conditions for $\bm{q}^{\gamma(0)}$, $\bm{q}^{\gamma(1)}$ and $\bm{q}^{\gamma(2)}$ are obtained
as follows:
For $ -p^2= P^2 \gg \Lambda^2$,
the photon matrix elements of the hadronic operators $O_n^i$
($i=S~ ({\rm or}\  \psi),G,NS$) can be calculated perturbatively.
Renormalising at $\mu^2 = P^2 $, we obtain at two-loop level
\begin{equation}
  \langle \gamma (p) \mid O_n^i (\mu) \mid \gamma (p) \rangle \vert_{\mu^2= P^2}
= \frac{\alpha}{4\pi} \Bigl\{ {\widetilde A}_{n}^{i(1)}  +\frac{\alpha_s(P^2)}{4\pi}
{\widetilde A}_{n}^{i(2)}  \Bigr\} ,\quad  i=S~ ({\rm or}\  \psi),G,NS~.
\label{Initial}
\end{equation}
The ${\widetilde A}_{n}^{i(1)}$ and
${\widetilde A}_{n}^{i(2)}$ terms represent the operator mixing between the
hadronic operators and photon operators in the NLO and NNLO, respectively, and
the operator mixing implies that there exist parton distributions in the
photon. Thus we have, at $\mu^2= P^2$ (or at $t=0$),
\begin{equation}
  \bm{q}^{\gamma(0)}(0)=\bm{0}, \qquad
 \bm{q}^{\gamma(1)}(0)= \frac{\alpha}{4\pi}  {\widetilde{\bm{A}}}_n^{(1)},\qquad
  \bm{q}^{\gamma(2)}(0)= \frac{\alpha\alpha_s(P^2)}{(4\pi)^2}
{\widetilde{\bm{A}}}_n^{(2)},
\end{equation}
with
\begin{equation}
{\widetilde{\bm{A}}}_n^{(l)}=\Bigl({\widetilde A}_{n}^{S(l)}, {\widetilde A}_{n}^{G(l)}, {\widetilde
A}_{n}^{NS(l)}   \Bigr)~, \qquad l=1, 2~.
\end{equation}
Actually, the initial quark distributions emerge in the NLO (the order $\alpha$) and
gluon distribution in the NNLO (the order $\alpha\alpha_s$).
The expressions of ${\widetilde{\bm{A}}}_n^{(1)}$ and ${\widetilde{\bm{A}}}_n^{(2)}$
in $\overline{\rm MS}$ scheme are enumerated in Appendix \ref{InitialConditions}.

With these initial conditions, the solutions ${\bm{q}}^{\gamma(0)}(t)$,
${\bm{q}}^{\gamma(1)}(t)$ and ${\bm{q}}^{\gamma(2)}(t)$ are given by
\bea
{\bm{q}}^{\gamma(0)}(t)&=&\frac{4\pi}{\alpha_s(t)}~ \bm{a}~
     \biggl\{{\bm{1}}-\biggl[\frac{\alpha_s(t)}{\alpha_s(0)}
   \biggr]^{1-\frac{2 P^{(0)}}{\beta_0}}  \biggr\}~, \label{SolLO} \\
 \nonumber  \\
{\bm{q}}^{\gamma(1)}(t)&=&
\biggl[\frac{\alpha}{2\pi}~{\bm{R}}_{K(1)} + 2\bm{a}~R_{P(1)}
\biggr]
\frac{-1}{P^{(0)}} ~\biggl\{{\bm{1}}-\biggl[\frac{\alpha_s(t)}{\alpha_s(0)}
   \biggr]^{-\frac{2 P^{(0)}}{\beta_0}} \biggr\}\nonumber\\
 &&-2~\bm{a}~
  \biggl\{  \int^t_0 d\tau  e^{( P^{(0)}-\frac{\beta_0}{2})\tau}~
R_{P(1)}~e^{- P^{(0)}\tau }
  \biggr\}~ e^{ P^{(0)}t}  +{\bm{q}}^{\gamma(1)}(0)\biggl[\frac{\alpha_s(t)}{\alpha_s(0)}
   \biggr]^{-\frac{2 P^{(0)}}{\beta_0}} ~, \nn\\\label{SolNLO}
\eea
\bea
{\bm{q}}^{\gamma(2)}(t)&=&\frac{\alpha_s(t)}{2\pi}\Biggl\{
\Bigl[\frac{\alpha}{2\pi}~{\bm{R}}_{K(2)} + 2{\bm{a}}~R_{P(2)} \Bigr]
\frac{-1}{\frac{\beta_0}{2}+P^{(0)}}
  \biggl\{{\bm{1}}-\biggl[\frac{\alpha_s(t)}{\alpha_s(0)}
   \biggr]^{-1-\frac{2 P^{(0)}}{\beta_0}}  \biggr\} \nonumber  \\
  & & \hspace{1.2cm}  -2{\bm{a}}
  \biggl\{  \int^t_0 d\tau
e^{(P^{(0)}-\frac{\beta_0}{2})\tau}~R_{P(2)}~e^{-(\frac{\beta_0}{2}+P^{(0)})\tau }
  \biggr\}e^{(\frac{\beta_0}{2}+P^{(0)})t} \nonumber \\
& & \hspace{1.2cm}+\Bigl[\frac{\alpha}{2\pi}~{{\bm{R}}}_{K(1)} + 2{\bm{a}}~R_{P(1)}
\Bigr] \frac{-1}{P^{(0)}} \nonumber\\
&&\hspace{4cm} \times\biggl\{  \int^t_0 d\tau   \Bigl[{\bm{1}} - e^{P^{(0)}\tau} \Bigr]R_{P(1)}
~e^{-(\frac{\beta_0}{2}+P^{(0)})\tau }\biggr\}e^{(\frac{\beta_0}{2}+P^{(0)})t}
\nonumber \\
  & & \hspace{1.2cm}  -2~{\bm{a}} \biggl\{  \int^t_0 d\tau
\biggl[   \int^\tau_0 d\tau'
e^{(P^{(0)}-\frac{\beta_0}{2})\tau'}~R_{P(1)}~e^{-P^{(0)}\tau' }
\biggr] \nonumber  \\
&&\hspace{4cm} \times e^{P^{(0)}\tau}R_{P(1)}~e^{-(\frac{\beta_0}{2}+P^{(0)})\tau }
\biggr\}e^{(\frac{\beta_0}{2}+P^{(0)})t}\nonumber \\
 & & \hspace{1.2cm} +{\bm{q}}^{\gamma(1)}(0) \biggl\{\int^t_0 d\tau  e^{P^{(0)}\tau}
R_{P(1)} ~e^{-(\frac{\beta_0}{2}+P^{(0)})\tau }\biggr\}e^{(\frac{\beta_0}{2}+P^{(0)})t}
\Biggr\} \nonumber\\
&& \hspace{1.2cm}+  {\bm{q}}^{\gamma(2)}(0)~\biggl[\frac{\alpha_s(t)}{\alpha_s(0)}
   \biggr]^{-\frac{2 P^{(0)}}{\beta_0}} , \label{SolNNLO}
\eea
where
\begin{equation}
  \bm{a}=\frac{\alpha}{2\pi\beta_0} \bm{k}^{(0)}
\frac{1}{1-\frac{2 P^{(0)}}{\beta_0}} ~.
\end{equation}
For the case of the real photon target, the NLO solution ${\bm{q}}^{\gamma(1)}(t)$ was given in \cite{GR} (see also \cite{FP2}) and the NNLO solution ${\bm{q}}^{\gamma(2)}(t)$
in \cite{MVV}.

The moments of the splitting functions are related to the anomalous dimensions
of operators as follows \cite{USU}:
\bea
    P^{(0)}&=&-\frac{1}{4}\widehat \gamma^{(0)}_n~, \qquad
    P^{(1)}=-\frac{1}{8}\widehat \gamma^{(1)}_n~, \qquad
    P^{(2)}=-\frac{1}{16}\widehat \gamma^{(2)}_n ~, \nonumber\\
   {\bm{k}}^{(0)}&=&\frac{1}{4}{\bm{K}}^{(0)}_n, \qquad
    {\bm{k}}^{(1)}=\frac{1}{8}{\bm{K}}^{(1)}_n, \qquad
    {\bm{k}}^{(2)}=\frac{1}{16}{\bm{K}}^{(2)}_n~. \label{SplittingVSAnoDimension}
\eea
The evaluation of ${\bm{q}}^{\gamma(0)}(t)$, ${\bm{q}}^{\gamma(1)}(t)$ and
${\bm{q}}^{\gamma(2)}(t)$
in Eqs.(\ref{SolLO})-(\ref{SolNNLO}) can be easily done by introducing
the projection operators $P^n_i$~\cite{BB}
\bea
      P^{(0)}&=&-\frac{1}{4}\widehat \gamma^{(0)}_n
=-\frac{1}{4}\sum_{i=+,-,NS}
\lambda^n_i~P^n_i,     \qquad   i=+,-,NS,  \label{EigenProjector}\\
   P^n_i~P^n_j&=&\begin{cases}
0 &\text{for}\quad i\ne j, \cr
                        P^n_i &\text{for}\quad i=j,   \qquad
\end{cases}
 \qquad  \qquad  \qquad   \sum_i  P^n_i ={\bf 1}~, \label{Projector}
\eea
where $\lambda^n_i$ are the eigenvalues of the matrix $\widehat \gamma^{(0)}_n$.
Then, rewriting $\alpha_s(0)$ and $\alpha_s(t)$ as $\alpha_s(P^2)$ and $\alpha_s(Q^2)$, respectively,
we obtain
\bea
  {\bm{q}}^{\gamma(0)}(t)/\Bigl[ \frac{\alpha}{8\pi \beta_0}\Bigr] &=&
   \frac{4\pi}{\alpha_s(Q^2)}~{\bm{K}}^{(0)}_n~\sum_i P^n_i~
       \frac{1}{1+d^n_i}
   \biggl\{1-\biggl[\frac{\alpha_s(Q^2)}{\alpha_s(P^2)}
   \biggr]^{1+d^n_i}  \biggr\}~,  \label{qv{(0)}(t)} \\
&&\nonumber  \\
 {\bm{q}}^{\gamma(1)}(t)/\Bigl[ \frac{\alpha}{8\pi \beta_0}\Bigr] &=&
\Biggl\{{\bm{K}}^{(1)}_n~\sum_i P^n_i~
\frac{1}{d^n_i} + \frac{\beta_1}{\beta_0}{\bm{K}}^{(0)}_n
~\sum_i P^n_i~\Bigl( 1 - \frac{1}{d^n_i} \Bigr)  \nonumber  \\
& &\quad - {\bm{K}}^{(0)}_n \sum_{j,i} \frac{P^n_j~ \widehat \gamma^{(1)}_n~P^n_i}
  {2\beta_0+\lambda^n_j-\lambda^n_i}~\frac{1}{d^n_i}~
   - 2\beta_0{\widetilde{\bm{A}}}_n^{(1)}~\sum_i P^n_i  \Biggr\} \nonumber  \\
& & \qquad \qquad \qquad \qquad \qquad \qquad \qquad \times
\biggl\{1-\biggl[\frac{\alpha_s(Q^2)}{\alpha_s(P^2)}
   \biggr]^{d^n_i}  \biggr\}\nonumber  \\
&+&  \Biggl\{{\bm{K}}^{(0)}_n\sum_{i,j}~\frac{ P^n_i~\widehat \gamma^{(1)}_n
~P^n_j}{2\beta_0+\lambda^n_i-\lambda_j^n}~
\frac{1}{1+d^n_i}
 - \frac{\beta_1}{\beta_0}{\bm{K}}^{(0)}_n~
\sum_i~P^n_i~\frac{d^n_i}{1+d^n_i}
\Biggr\} \nonumber  \\
& & \qquad \qquad \qquad \qquad \qquad \qquad \qquad \times
\biggl\{1-\biggl[\frac{\alpha_s(Q^2)}{\alpha_s(P^2)}
   \biggr]^{1+d^n_i}  \biggr\} \nonumber  \\
  &+& 2\beta_0{\widetilde{\bm{A}}}_n^{(1)}~, \label{qv{(1)}(t)}
\eea
%%%%%%%%%%%%%%%%%%%%%%%%%%
\bea
{\bm{q}}^{\gamma(2)}(t)/\Bigl[\frac{\alpha}{8\pi\beta_0}
\Bigr]\Bigl[\frac{\alpha_s(Q^2)}{4\pi}\Bigr]&& \nonumber\\
&&\hspace{-3.5cm}=
\Biggl\{-\bm{K}^{(0)}_n\Bigl(\frac{\beta_1}{\beta_0}\Bigr)^2\sum_i
P^n_i~\Bigl(1- \frac{d_i^n}{2} \Bigr)+\bm{K}^{(0)}_n\frac{\beta_2}{\beta_0}\sum_i
P^n_i~\frac{1}{1-d_i^n}\Bigl(1- \frac{d_i^n}{2} \Bigr)
\nonumber\\
&&\hspace{-2.8cm}
-\bm{K}^{(0)}_n\frac{\beta_1}{\beta_0}\biggl[\sum_{j,i}~\frac{P^n_j~\widehat\gamma^{(1)}_n~
P^n_i}{2\beta_0+\lambda_j^n-\lambda_i^n}~
\frac{1-d_j^n}{1-d_i^n}
+\sum_{j,i}~\frac{P^n_j~\widehat\gamma^{(1)}_n~
P^n_i}{4\beta_0+\lambda_j^n-\lambda_i^n}~
\frac{1-d_i^n+d_j^n}{1-d_i^n}\biggr]\nonumber\\
&&\hspace{-2.8cm}+\bm{K}^{(0)}_n\sum_{j,i}~\frac{P^n_j~
\widehat\gamma^{(2)}_nP^n_i}{4\beta_0+\lambda_j^n-\lambda_i^n}
~\frac{1}{1-d_i^n} \nonumber\\
&&\hspace{-2.8cm}
-\bm{K}^{(0)}_n\sum_{j,k,i} ~
\frac{P^n_j~\widehat \gamma^{(1)}_n~
P^n_k ~\widehat \gamma^{(1)}_n~P^n_i}
{(2\beta_0-\lambda_i^n+\lambda_k^n)(4\beta_0+\lambda_j^n-\lambda_i^n)}~
\frac{1}{1-d_i^n}
\nonumber\\
&&\hspace{-2.8cm}+\bm{K}^{(1)}_n\frac{\beta_1}{\beta_0}\sum_i P^n_i
+\bm{K}^{(1)}_n\sum_{j,i}~\frac{P^n_j~
\widehat\gamma^{(1)}_nP^n_i}{2\beta_0+\lambda_j^n-\lambda_i^n}
~\frac{1}{1-d_i^n} -\bm{K}^{(2)}_n\sum_i P^n_i~\frac{1}{1-d_i^n} \nonumber\\
&&\hspace{-2.8cm}+2\beta_0{\widetilde {\bm{A}}}_n^{(1)}
\sum_{j,i}~\frac{P^n_j~\widehat\gamma^{(1)}_nP^n_i}{2\beta_0+\lambda_j^n-\lambda_i^n}
-2\beta_0{\widetilde {\bm{A}}}_n^{(1)}\frac{\beta_1}{\beta_0}\sum_i P^n_id_i^n
-2\beta_0{\widetilde {\bm{A}}}_n^{(2)}\sum_i P^n_i \Biggr\}\nonumber\\
&&\hspace{4cm}\times\biggl\{1- \Bigl[\frac{\alpha_s(Q^2)}{\alpha_s(P^2)}
   \Bigr]^{d^n_i-1}  \biggr\}\nonumber \\
%%%%%%%%%%%%%%%%%%
&&\hspace{-3.4cm}+\Biggl\{\bm{K}^{(0)}_n\Bigl(\frac{\beta_1}{\beta_0}\Bigr)^2\sum_i
P^n_i~(1- d_i^n)-\bm{K}^{(0)}_n\frac{\beta_1}{\beta_0}\sum_{i,j}~\frac{P^n_i
~\widehat\gamma^{(1)}_n~P^n_j}{2\beta_0+\lambda_i^n-\lambda_j^n}~
\frac{1-d_i^n}{d_i^n}
\nonumber\\
&&\hspace{-2.8cm}
+\bm{K}^{(0)}_n\frac{\beta_1}{\beta_0}\sum_{j,i}~\frac{P^n_j~\widehat\gamma^{(1)}_n~
P^n_i}{2\beta_0+\lambda_j^n-\lambda_i^n}
-\bm{K}^{(0)}_n\sum_{j,i,k} ~
\frac{P^n_j~\widehat \gamma^{(1)}_n~
P^n_i ~\widehat \gamma^{(1)}_n~P^n_k}
{(2\beta_0+\lambda_i^n-\lambda_k^n)(2\beta_0+\lambda_j^n-\lambda_i^n)}~
\frac{1}{d_i^n}
\nonumber\\
&&\hspace{-2.8cm}
\nonumber\\
&&\hspace{-2.8cm}-\bm{K}^{(1)}_n\frac{\beta_1}{\beta_0}\sum_i P^n_i
+\bm{K}^{(1)}_n\sum_{i,j}~\frac{P^n_i~
\widehat\gamma^{(1)}_nP^n_j}{2\beta_0+\lambda_i^n-\lambda_j^n}
~\frac{1}{d_i^n}  \nonumber\\
&&\hspace{-2.8cm}-2\beta_0{\widetilde {\bm{A}}}_n^{(1)}
\sum_{i,j}~\frac{P^n_i~\widehat\gamma^{(1)}_nP^n_j}{2\beta_0+\lambda_i^n-\lambda_j^n}
+2\beta_0{\widetilde {\bm{A}}}_n^{(1)}\frac{\beta_1}{\beta_0}\sum_i P^n_id_i^n
\Biggr\} \biggl\{1- \Bigl[\frac{\alpha_s(Q^2)}{\alpha_s(P^2)}
   \Bigr]^{d^n_i}  \biggr\} \nonumber  \\
%%%%%%%%%%%%%%%%%%%%
 &&\hspace{-3.4cm}+\Biggl\{\bm{K}^{(0)}_n\Bigl(\frac{\beta_1}{\beta_0}\Bigr)^2\sum_i
P^n_i~\frac{d_i^n}{2} -\bm{K}^{(0)}_n\frac{\beta_2}{\beta_0}\sum_i
P^n_i~\frac{d_i^n}{2(1+d_i^n)}
\nonumber\\
&&\hspace{-2.8cm}
-\bm{K}^{(0)}_n\frac{\beta_1}{\beta_0}\biggl[\sum_{i,j}~\frac{P^n_i~\widehat\gamma^{(1)}_n~
P^n_j}{2\beta_0+\lambda_i^n-\lambda_j^n}~
\frac{d_j^n}{1+d_i^n}
+\sum_{i,j}~\frac{P^n_i~\widehat\gamma^{(1)}_n~
P^n_j}{4\beta_0+\lambda_i^n-\lambda_j^n}~
\frac{1+d_i^n-d_j^n}{1+d_i^n}\biggr]\nonumber\\
&&\hspace{-2.8cm}+\bm{K}^{(0)}_n\sum_{i,j}~\frac{P^n_i~
\widehat\gamma^{(2)}_nP^n_j}{4\beta_0+\lambda_i^n-\lambda_j^n}
~\frac{1}{1+d_i^n} \nonumber\\
&&\hspace{-2.8cm}
+\bm{K}^{(0)}_n\sum_{i,j,k} ~
\frac{P^n_i~\widehat \gamma^{(1)}_n~
P^n_j ~\widehat \gamma^{(1)}_n~P^n_k}
{(2\beta_0+\lambda_i^n-\lambda_j^n)(4\beta_0+\lambda_i^n-\lambda_k^n)}~
\frac{1}{1+d_i^n}
\Biggr\}
\biggl\{1- \Bigl[\frac{\alpha_s(Q^2)}{\alpha_s(P^2)}
   \Bigr]^{d^n_i+1}  \biggr\}\nonumber \\
&&\nonumber \\
&&\hspace{-3.4cm}+2\beta_0{\widetilde {\bm{A}}}_n^{(2)} ~,\label{qv{(2)}(t)}
\eea
where $d^n_i\equiv\frac{\lambda^n_i}{2\beta_0}$.
The information on the parameters is given in Appendix A. See also Ref.~\cite{USU}.

Finally, since
$\bm{q}^\gamma(t)=\bm{q}^{\gamma(0)}(t)+\bm{q}^{\gamma(1)}(t)
+\bm{q}^{\gamma(2)}(t)$, and from (\ref{PartonRowVector}), the moments of
the flavour-singlet quark, gluon and flavour-nonsinglet quark distributions
in the virtual photon are given, respectively, by
\bea
 q_S^\gamma(n,Q^2,P^2)=(1,1)\ {\rm component\ of\ the\  row\ vector\ }\bm{q}^\gamma(t)~, \\
 G^\gamma(n,Q^2,P^2)=(1,2)\ {\rm component\ of\ the\  row\ vector\ }\bm{q}^\gamma(t)~, \\
 q_{NS}^\gamma(n,Q^2,P^2)=(1,3)\ {\rm component\ of\ the\  row\ vector\ }\bm{q}^\gamma(t)~.
\eea

For $n\!=\!2$, one of the eigenvalues, $\lambda_-^{n=2}$, in Eq.(\ref{EigenProjector})
vanishes and we have $d_-^{n=2}\!=\!0$.
This is due to the fact that the corresponding operator is the hadronic
energy-momentum tensor and is, therefore, conserved with a null anomalous dimension~\cite{BB}.
The second moments of the singlet quark and gluon distributions, $q_S^\gamma(n\!=\!2,Q^2,P^2)$ and
$G^\gamma(n\!=\!2,Q^2,P^2)$, have terms which are proportional to
$\frac{1}{d_-^{n=2}}$ and thus diverge. However, we see from (\ref{qv{(1)}(t)}) and (\ref{qv{(2)}(t)}) that
these terms are multiplied by a factor
$\Bigl[1-\Bigl(\alpha_s(Q^2)/\alpha_s(P^2)\Bigr)^{d_-^{n=2}}\Bigr]$ which vanishes. In the end, the
second moments $q_S^\gamma(n\!=\!2,Q^2,P^2)$ and
$G^\gamma(n\!=\!2,Q^2,P^2)$ remain finite~\cite{UW2}.

\section{Renormalisation scheme dependence \label{RenormSchemeDependence}}

The structure functions of the photon (nucleon) are expressed as convolutions of
coefficient functions and parton distributions of the target photon (nucleon).
But it is well known that these coefficient functions and parton distributions are
by themselves renormalisation-scheme
dependent. There are two kinds of renormalisation-scheme dependence:
(i) One is the dependence on the renormalisation-prescription (RP) chosen
 in defining the QCD coupling constant $\alpha_s$.
(ii) The other is the so-called factorisation-scheme (FS) dependence.
The coefficient functions and parton distributions (equivalently, the anomalous dimensions
and photon matrix elements of operators) are dependent on the
FS adopted for defining these quantities.
Of course, the physically measurable quantities such as structure functions are
independent of the choice of the RP for $\alpha_s$
and also of the choice of the FS.

In this section we will show first that the parton
distributions of the virtual photon up to the NNLO, which were obtained in
Eqs.(\ref{qv{(0)}(t)})-(\ref{qv{(2)}(t)}), are independent of the RP chosen to define $\alpha_s$. Then we consider the parton distributions in the
virtual photon target in the two factorisation schemes, $\overline {\rm MS}$ and
${\rm DIS}_{\gamma}$ schemes.

\subsection{Independence of the renormalisation prescription for $\alpha_s$}
We show that the parton distributions of the virtual photon up to the NNLO are independent of the
choice of the RS in defining the QCD coupling constant $\alpha_s$.
The beta function $\beta(g)$ is expanded in powers of $g^2$ up to the three-loop level as
\begin{equation}
\beta(g)=-\frac{g^3}{16\pi^2}\beta_0-\frac{g^5}{(16\pi^2)^2}\beta_1
-\frac{g^7}{(16\pi^2)^3}\beta_2 + \cdots~.
\label{Beta}
\end{equation}
Then the QCD running coupling constant $\alpha_s(Q^2)$ is expressed as~\cite{ParticleData},
\begin{equation}
    \frac{\alpha_s(Q^2)}{4\pi}
    = \frac{1}{\beta_0 L}
      - \frac{1}{(\beta_0 L)^2} \frac{\beta_1}{\beta_0} \ln L
      + \frac{1}{(\beta_0 L)^3} \left(\frac{\beta_1}{\beta_0}\right)^2
            \left[ \left(\ln L - \frac{1}{2} \right)^2
            + \frac{\beta_0 \beta_2}{\beta_1^2} - \frac{5}{4} \right]
      + {\cal O} \left(\frac{1}{L^4}\right) , \label{runningAlpha}
\end{equation}
where $L\!=\!\ln(Q^2/\Lambda^2)$.
It is known that the first two coefficients $\beta_0$ and $\beta_1$ in (\ref{Beta})
are renormalisation prescription independent but the coefficient $\beta_2$ is not~\cite{Muta}.

Suppose that $\beta_2$ is obtained in one scheme, for example, in the momentum-space subtraction (MOM)
scheme~\cite{CelGon}, and
let $\delta \beta_2$ be a difference between $\beta_2|_{\rm one\ scheme}$
and the one calculated in the modified minimal subtraction ($\overline{\rm MS}$) scheme~\cite{BBDM},
\begin{equation}
\delta \beta_2=\beta_2|_{\rm one\ scheme}-\beta_2|_{\overline{\rm MS}}~.
\end{equation}
Then we find from (\ref{runningAlpha}) that the change in the renormalisation prescription for $\alpha_s$
(in other words, the change $\delta \beta_2$)
has an effect on the running coupling constant $\alpha_s(Q^2)$ as
\begin{equation}
\delta\Bigl(\frac{ \alpha_s(Q^2)}{4\pi}\Bigr)= \frac{1}{(\beta_0 L)^3} \frac{1}{\beta_0}\delta \beta_2
+{\cal O} \Bigl(\frac{1}{L^4}\Bigr)
= \left[\frac{\alpha_s(Q^2)}{4\pi}   \right]^3 \frac{1}{\beta_0}\delta \beta_2+
{\cal O} (\alpha_s^4)~,
\label{ChangeofALPHA}
\end{equation}
so that the effect on the order $\alpha_s^N$ appears in the higher orders of $\alpha_s^{N+2}$.

The change in the renormalisation prescription for $\alpha_s$ leads to the change of the coefficient functions,
the anomalous dimensions and photon matrix elements of hadronic operators.
We see from (\ref{qv{(0)}(t)})-(\ref{qv{(2)}(t)}) that the parton distributions in the photon
up to the NNLO are expressed in terms of the
anomalous dimensions of the hadronic operators calculated up to the order
$\alpha_s^3$ (the three-loop level), the mixing anomalous dimensions between the photon and hadronic
operators up to the order $\alpha\alpha_s^2$ (the three-loop level), and
photon matrix elements up to the order $\alpha\alpha_s$ (the two-loop level). Therefore, only relevant
is the three-loop anomalous dimension matrix in the hadronic sector. To see this, we use
(\ref{ChangeofALPHA}) and we obtain,
\bea
 {\widehat \gamma}_{n}(\alpha'_s)
&=&\frac{\alpha'_s}{4\pi}{\widehat \gamma}_{n}^{'(0)}+
\frac{{\alpha'}^2_s}{(4\pi)^2}{\widehat \gamma}_{n}^{'(1)} +
\frac{{\alpha'}^3_s}{(4\pi)^3}{\widehat \gamma}_{n}^{'(2)}+ \cdots~
 \nonumber \\
&=&\Bigl\{\frac{\alpha_s}{4\pi} + \Bigl(\frac{\alpha_s}{4\pi}\Bigr)^3
\frac{\delta \beta_2}{\beta_0} \Bigr\} {\widehat \gamma}_{n}^{'(0)}+
\frac{\alpha^2_s}{(4\pi)^2}{\widehat \gamma}_{n}^{'(1)} +
\frac{\alpha^3_s}{(4\pi)^3}{\widehat \gamma}_{n}^{'(2)}+ \cdots~\nonumber \\
&=&\frac{\alpha_s}{4\pi}{\widehat \gamma}_{n}^{(0)}+
\frac{\alpha^2_s}{(4\pi)^2}{\widehat \gamma}_{n}^{(1)} +
\frac{\alpha^3_s}{(4\pi)^3}{\widehat \gamma}_{n}^{(2)}+ \cdots~.
\label{GammaExpansion}
\eea
Thus we get
\bea
{\widehat \gamma}_{n}^{'(0)}&=&{\widehat \gamma}_{n}^{(0)}~,
\quad {\widehat \gamma}_{n}^{'(1)}={\widehat \gamma}_{n}^{(1)}~,
\qquad \delta {\widehat \gamma}_{n}^{(2)}\equiv {\widehat \gamma}_{n}^{'(2)}-{\widehat \gamma}_{n}^{(2)}
=-{\widehat \gamma}_{n}^{(0)}\frac{\delta \beta_2}{\beta_0}~. \label{deltagamma2}
\eea

Let us write the difference of the parton distributions in the photon between
one scheme chosen to define $\alpha_s$ and $\overline{\rm MS}$ scheme as
\begin{equation}
\delta{\bm{q}}^{\gamma(i)}(t)~,\qquad \qquad i=0,1,2,
\end{equation}
where $i=0,1,2$ denote the LO, NLO and NNLO expressions, respectively.
Now we find the difference in the LO expression ${\bm{q}}^{\gamma(0)}(t)$ given in (\ref{qv{(0)}(t)}) is
\bea
\delta{\bm{q}}^{\gamma(0)}(t)/\Bigl[ \frac{\alpha}{8\pi \beta_0}\Bigr] &=&
 \delta \Bigl(\frac{4\pi}{\alpha_s(Q^2)} \Bigr)~{\bm{K}}^{(0)}_n~\sum_i P^n_i~
       \frac{1}{1+d^n_i}
   \biggl\{1-\biggl[\frac{\alpha_s(Q^2)}{\alpha_s(P^2)}
   \biggr]^{1+d^n_i}  \biggr\} \nonumber \\
&&+\frac{4\pi}{\alpha_s(Q^2)}~{\bm{K}}^{(0)}_n~\sum_i P^n_i~
       \frac{1}{1+d^n_i}
 \biggl\{-\delta\biggl(\biggl[\frac{\alpha_s(Q^2)}{\alpha_s(P^2)}
   \biggr]^{1+d^n_i} \biggr) \biggr\} \nonumber \\
&=&\frac{\alpha_s(Q^2)}{4\pi}\frac{\delta
\beta_2}{\beta_0}\bm{K}_n^{(0)}\sum_{i}P_i^n
\Biggl(\frac{d_i^n}{d_i^n+1}\biggl\{1-\biggl[\frac{\alpha_s(Q^2)}{\alpha_s(P^2)}\biggr]^{d_i^n+1}
\biggr\} \nonumber\\
&&\hspace{6cm}-\biggl\{1-\biggl[\frac{\alpha_s(Q^2)}{\alpha_s(P^2)}\biggr]^{d_i^n-1}
\biggr\}  \Biggr)~,
 \label{deltaqv{(0)}(t)}
\eea
where we have used the following formulae:
\bea
\delta \Bigl(\frac{4\pi}{\alpha_s(Q^2)} \Bigr)&=&-\frac{\alpha_s(Q^2)}{4\pi} \frac{1}{\beta_0}\delta \beta_2~,
\label{formuladelta1}\\
\delta \left(\left[\frac{\alpha_s(Q^2)}{\alpha_s(P^2)}\right]^a \right)&=&
\left[\frac{\alpha_s(Q^2)}{\alpha_s(P^2)} \right]^a~\left\{
\Bigl(\frac{\alpha_s(Q^2)}{4\pi}\Bigr)^2-\Bigl(\frac{\alpha_s(P^2)}{4\pi}\Bigr)^2
\right\}\frac{a}{\beta_0}\delta\beta_2~. \label{formuladelta2}
\eea
On the other hand, we see that $\beta_2$ and ${\widehat \gamma}_{n}^{(2)}$ do not appear in the NLO expression
${\bm{q}}^{\gamma(1)}(t)$ given in (\ref{qv{(1)}(t)}). Also we already know the effect of the change in
$\beta_2$ on the order $\alpha_s^N$ appears in the higher orders of $\alpha_s^{N+2}$. Thus,
as far as the analysis for the parton distributions up to the NNLO is concerned, we conclude
\begin{equation}
\delta{\bm{q}}^{\gamma(1)}(t)=0~. \label{deltaqv{(1)}(t)}
\end{equation}
Finally, we notice that $\beta_2$ and ${\widehat \gamma}_{n}^{(2)}$ appear in the NNLO expression of
${\bm{q}}^{\gamma(2)}(t)$ given in (\ref{qv{(2)}(t)}).
Therefore, we get
\bea
\delta{\bm{q}}^{\gamma(2)}(t)/\Bigl[ \frac{\alpha}{8\pi \beta_0}\Bigr]
\Bigl[\frac{\alpha_s(Q^2)}{4\pi}\Bigr]&&\nonumber\\
&&\hspace{-3.5cm}=
\biggl\{\bm{K}^{(0)}_n\frac{\delta\beta_2}{\beta_0}\sum_i
P^n_i~\frac{1}{1-d_i^n}\Bigl(1- \frac{d_i^n}{2} \Bigr)\nonumber\\
&&\hspace{-1.5cm}+ \bm{K}^{(0)}_n\sum_{j,i}~\frac{P^n_j~
\delta\widehat\gamma^{(2)}_nP^n_i}{4\beta_0+\lambda_j^n-\lambda_i^n}
~\frac{1}{1-d_i^n} \biggr\}
   \biggl\{1-\biggl[\frac{\alpha_s(Q^2)}{\alpha_s(P^2)}
   \biggr]^{d^n_i-1}  \biggr\} \nonumber \\
&&\hspace{-3.3cm}+\biggl\{-\bm{K}^{(0)}_n\frac{\delta\beta_2}{\beta_0}\sum_i
P^n_i~\frac{d_i^n}{2(1+d_i^n)}\nonumber\\
 &&\hspace{-1.5cm}+\bm{K}^{(0)}_n\sum_{i,j}~\frac{P^n_i~
\delta\widehat\gamma^{(2)}_nP^n_j}{4\beta_0+\lambda_i^n-\lambda_j^n}
~\frac{1}{1+d_i^n}\biggr\}
   \biggl\{1-\biggl[\frac{\alpha_s(Q^2)}{\alpha_s(P^2)}
   \biggr]^{d^n_i+1}  \biggr\} ~.
\eea
%%%
Now Eq.(\ref{deltagamma2}) and the properties of the projection operators in (\ref{Projector}) give
\begin{equation}
\sum_{j}~\frac{P^n_j~
\delta\widehat\gamma^{(2)}_nP^n_i}{4\beta_0+\lambda_j^n-\lambda_i^n}=
\sum_{j}~\frac{P^n_i~
\delta\widehat\gamma^{(2)}_nP^n_j}{4\beta_0+\lambda_i^n-\lambda_j^n}=-P^n_i\frac{\lambda_i^n}{4\beta_0}
\frac{\delta\beta_2}{\beta_0}~,
\end{equation}
and we obtain
\bea
\delta{\bm{q}}^{\gamma(2)}(t)/\Bigl[ \frac{\alpha}{8\pi
\beta_0}\Bigr]
&=&\Bigl[\frac{\alpha_s(Q^2)}{4\pi}\Bigr]\frac{\delta\beta_2}{\beta_0}
\Biggl(\bm{K}^{(0)}_n\sum_i P^n_i~
   \biggl\{1-\biggl[\frac{\alpha_s(Q^2)}{\alpha_s(P^2)}
   \biggr]^{d^n_i-1}  \biggr\} \nonumber \\
&&\qquad \qquad  -\bm{K}^{(0)}_n\sum_i
P^n_i~\frac{d_i^n}{(1+d_i^n)}
   \biggl\{1-\biggl[\frac{\alpha_s(Q^2)}{\alpha_s(P^2)}
   \biggr]^{d^n_i+1}  \biggr\}\Biggr)
% \nonumber \\
~,
 \label{deltaqv{(2)}(t)}
\eea
which exactly cancels the change $\delta{\bm{q}}^{\gamma(0)}(t)$ given in (\ref{deltaqv{(0)}(t)}).
Thus we find
\begin{equation}
\delta{\bm{q}}^{\gamma(0)}(t)+\delta{\bm{q}}^{\gamma(1)}(t)+\delta{\bm{q}}^{\gamma(2)}(t)=0~.
\end{equation}
The parton distributions in the photon up to the NNLO are, indeed, independent of the
renormalisation-prescription adopted to define the QCD coupling constant $\alpha_s$.

%Finally we remark that we can show in the same way that the parton distributions in the nucleon are
%independent of the renormalisation-prescription for $\alpha_s$.

%%%%%%%%%%%%%%%

%%%%%%%%%%%%%%%%%  Factorisation schemes %%%%%%%%%%%%%%%%%%%
\subsection{Factorisation schemes}
%%%%%%%%%%%%%%%%%%%%%%%%%%%%%%%%%%%%%%%%%%%%%%%%%%%%%%%%%%%%%%

The structure functions are expressed as convolutions of parton distributions and coefficient
functions. The Mellin moments of the virtual photon structure function $\frac{1}{x}F_2^{\gamma}(x,Q^2,P^2)$ is expressed as
\bea
     \int_0^1 dx x^{n-1}\frac{1}{x}~  F_2^{\gamma}(x,Q^2,P^2)
&\equiv&F_2^{\gamma}(n,Q^2,P^2)  \nonumber \\
&=& \widetilde{\bm{q}}^{\gamma}(n,Q^2,P^2)\cdot \widetilde{\bm{C}}_2(n,Q^2)
\label{F2moment}~,
\eea
where $\widetilde{\bm{q}}^\gamma(n,Q^2,P^2)$ is a four-component row vector
\begin{equation}
\widetilde{\bm{q}}^\gamma (n,Q^2,P^2)\equiv\Bigl(q^{\gamma}_S (n,Q^2,P^2), G^{\gamma}
(n,Q^2,P^2),q^{\gamma}_{NS}(n,Q^2,P^2),
\Gamma_n^{\gamma}\Bigr) ~,
\end{equation}
with $\Gamma_n^{\gamma}=1$, the moment of the photon distribution function
(see Eq.(\ref{PhotonDistribution})), being added to the row vector ${\bm{q}}^\gamma$ of
(\ref{PartonRowVector}),
 and
$\widetilde{\bm{C}}_2(n,Q^2)$ is a four component column vector
\bea
   \widetilde{\bm{C}}_2(n,Q^2)&\equiv&( C_2^S(n,Q^2),~
  C_2^G(n,Q^2),~C_2^{NS}(n,Q^2),~
   C_2^\gamma(n,Q^2))^{\rm T} \label{Coefficient} \\
&=&({\bm{C}}_2(n,Q^2),~ C_2^\gamma(n,Q^2))^{\rm T}~, \nonumber
\eea
where the hadronic coefficient functions ${\bm{C}}_2=(C_2^S, C_2^G, C_2^{NS})$
are made up of the flavour-singlet quark, gluon and nonsinglet quark, and
$C_2^\gamma$ is the photonic coefficient functions.

Since $F_2^{\gamma}$ is a physical quantity,
its Mellin moments $F_2^{\gamma}(n,Q^2,P^2)$ in (\ref{F2moment}) is unique and FS-independent. But
there remains a freedom in the factorisation of
$F_2^{\gamma}$ into $\widetilde{\bm{q}}^{\gamma}$ and $\widetilde{\bm{C}}_2$.
Given the formula (\ref{F2moment}),
we can always redefine $\widetilde{\bm{q}}^{\gamma}$ and $\widetilde{\bm{C}}_2$ as
follows:
\bea
 \widetilde{\bm{q}}^{\gamma}(n,Q^2,P^2)  &\rightarrow& \widetilde{\bm{q}}^{\gamma}(n,Q^2,P^2)\vert_a
 \equiv\widetilde{\bm{q}}^{\gamma}(n,Q^2,P^2)~Z_a(n,Q^2)~, \label{qaVsqMSbar}\\
\widetilde{\bm{C}}_2(n,Q^2)&\rightarrow& \widetilde{\bm{C}}_2(n,Q^2)\vert_a
      \equiv Z^{-1}_a(n,Q^2)~\widetilde{\bm{C}}_2(n,Q^2)~,
\label{CaVsCMSbar}
\eea
where $
\widetilde{\bm{q}}(n,Q^2,P^2)\vert_a $ and $\widetilde{\bm{C}}_2(n,Q^2)\vert_a$ correspond to the
quantities in a new factorisation scheme-$a$. It is noted that coefficient functions and
anomalous dimensions are closely connected under factorisation.
In the following we will study
the parton distributions in the virtual photon target up to the NNLO in two factorisation schemes,
namely, $\overline {\rm MS}$ and ${\rm DIS}_{\gamma}$ schemes.
%%%%%%%%%%%

\subsubsection{The $\overline {\rm MS}$ scheme}

This is the only scheme in which the relevant quantities, namely,
the $\beta$ function parameters up to three-loop level~\cite{TVZ},
anomalous dimensions up to
three-loop level~\cite{MVV1, MVV2, MVV3} and photon matrix elements up to two-loop level~\cite{MSvN}, were actually calculated.
We insert them into the formulae
given by Eqs.(\ref{qv{(0)}(t)})-(\ref{qv{(2)}(t)}) and obtain the moments of the parton distributions predicted by $\overline {\rm MS}$ scheme.

There is another way, a simpler one indeed,
to obtain $\bm{q}^\gamma(n,Q^2,P^2)|_{\overline{\rm MS}}$ up to the NNLO,
once we know the expression for the moment sum rule of $F_2^\gamma(x,Q^2,P^2)$
obtained up to the NNLO and all the quantities in the expression
are the ones calculated in $\overline {\rm MS}$ scheme.
We will show it in Appendix \ref{MomentF_2}.

\subsubsection{The ${\rm DIS}_{\gamma}$ scheme}
An interesting factorisation scheme, which is called ${\rm
DIS}_{\gamma}$, was introduced some time ago into the NLO analysis of the
unpolarised real photon structure function $F_2^{\gamma}(x, Q^2)$.
Gl{\"u}ck, Reya and Vogt~\cite{GRV} observed that, in $\overline {\rm MS}$
scheme, the ${\rm ln}(1-x)$ term in the one-loop photonic coefficient function
$C_2^{\gamma}(x)$ for $F_2^{\gamma}$, which becomes negative and divergent for
$x \rightarrow 1$, drives the `pointlike' part of $F_2^{\gamma}$ to large
negative values as $x \rightarrow 1$, leading to a strong difference
between the LO and the NLO results for
$F_{2,{\rm pointlike}}^{\gamma}$ in the large-$x$ region.
They introduced the ${\rm DIS}_{\gamma}$ scheme in which
the photonic coefficient function $C_2^{\gamma}$, i.e., the direct-photon
contribution to $F_2^{\gamma}$, is absorbed into the photonic quark
distributions.
A similar situation occurs in the polarised case, and the ${\rm DIS}_{\gamma}$
scheme was applied to the NLO analysis for the
spin-dependent structure function $g_1^{\gamma}(x, Q^2)$ of the real photon
target~\cite{SV}.

The transformation rule from $\overline {\rm MS}$ scheme to
${\rm DIS}_{\gamma}$ scheme was derived up to the NNLO in Ref.\cite{MVV}.
The moments of the parton distributions in ${\rm DIS}_{\gamma}$ scheme
are obtained as follows. In this scheme, the hadronic coefficient functions
are the same as their counterparts in ${\overline {\rm MS}}$ scheme,
but the photonic coefficient function is absorbed into the quark
distributions and thus set to zero,
\begin{equation}
{\bm{C}}_{2}(n,Q^2)|_{{\rm DIS}_{\gamma}}={\bm{C}}_{2}(n,Q^2)|_{\overline {\rm MS}}~,
\qquad C_{2}^{\gamma}(n,Q^2)|_{{\rm DIS}_{\gamma}}=0.
\end{equation}
Then Eq.(\ref{F2moment}) gives
\bea
F_2^{\gamma}(n,Q^2,P^2)
&=&\bm{q}^{\gamma}(n,Q^2,P^2)|_{{\rm DIS}_{\gamma}} \cdot \bm{C}_2(n,Q^2)|_{{\rm DIS}_{\gamma}}
\nonumber\\
&=&\bm{q}^{\gamma}(n,Q^2,P^2)|_{{\rm DIS}_{\gamma}}\cdot
\bm{C}_2(n,Q^2)|_{\overline{\rm MS}}~. \label{F2DISgamma}
\eea
On the other hand, $F_2^{\gamma}(n,Q^2,P^2)$ is expressed in ${\overline {\rm MS}}$ scheme as
\begin{equation}
F_2^{\gamma}(n,Q^2,P^2) =\bm{q}^{\gamma}(n,Q^2,P^2)|_{\overline{\rm MS}}\cdot
\bm{C}_2(n,Q^2)|_{\overline{\rm MS}}+ C_2^\gamma(n,Q^2)|_{\overline{\rm MS}}~.\label{F2MSbar}
\end{equation}

The expansion is made for $\bm{q}^{\gamma}(n,Q^2,P^2)|_{{\rm DIS}_{\gamma}}$
in terms of the LO, NLO and NNLO distributions as
\begin{equation}
\bm{q}^{\gamma}(n,Q^2,P^2)|_{{\rm DIS}_{\gamma}}=\bm{q}_n^{\gamma(0)}
+\bm{q}_n^{\gamma(1)}|_{{\rm DIS}_{\gamma}}+\bm{q}_n^{\gamma(2)}|_{{\rm DIS}_{\gamma}}+\cdots~, \label{PartonExpanded}
\end{equation}
where the LO $\bm{q}_n^{\gamma(0)}$ is FS-independent.
Denoting the difference of $\bm{q}_n^{\gamma(l)}|_{{\rm DIS}_{\gamma}}$ $(l=1,2)$
from ${\overline {\rm MS}}$ scheme predictions as $\delta{\bm{q}}_n^{\gamma(l)}|_{{\rm DIS}_{\gamma}}$, we write
\begin{equation}
{\bm{q}}_n^{\gamma(l)}|_{{\rm DIS}_{\gamma}}\equiv
{\bm{q}}_n^{\gamma(l)}|_{\overline {\rm MS}}+\delta{\bm{q}}_n^{\gamma(l)}|_{{\rm DIS}_{\gamma}} ~,
\qquad l=1, 2~. \label{SchemeDifference}
\end{equation}
Also the hadronic and photonic coefficient functions, $\bm{C}_2(n,Q^2)|_{\overline{\rm MS}}$ and $C_2^\gamma(n,Q^2)|_{\overline{\rm MS}}$, are expanded
in powers of $\alpha_s(Q^2)$ up to the NNLO as
\bea
\bm{C}_2(n,Q^2)|_{\overline{\rm MS}}&=&\bm{C}_{2,n}^{(0)}
+\frac{\alpha_s(Q^2)}{4\pi}\bm{C}_{2,n}^{(1)}|_{\overline{\rm MS}} +
\frac{\alpha^2_s(Q^2)}{(4\pi)^2}\bm{C}_{2,n}^{(2)}|_{\overline{\rm MS}}+\cdots~, \label{HadronicCoefficientExpanded} \\
C_2^\gamma(n,Q^2)|_{\overline{\rm MS}}&=&\frac{\alpha}{4\pi}3n_f \langle e^4 \rangle\Bigl\{
c_{2,n}^{\gamma(1)}|_{\overline{\rm MS}}
+\frac{\alpha_s(Q^2)}{4\pi}c_{2,n}^{\gamma(2)}|_{\overline{\rm MS}}+\cdots \Bigr\}~,
\label{PhotonicCoefficientExpanded}
\eea
where $\langle e^4\rangle=\sum_ie_i^4/n_f$ and $\bm{C}_{2,n}^{(0)}$ is FS-independent.
Now putting (\ref{PartonExpanded})-(\ref{HadronicCoefficientExpanded})
into the r.h.s. of (\ref{F2DISgamma}) and comparing the result
with (\ref{F2MSbar}) and (\ref{PhotonicCoefficientExpanded}), the following relations
are obtained,
\bea
\frac{\alpha}{4\pi}3n_f \langle e^4 \rangle c_{2,n}^{\gamma(1)}|_{\overline{\rm MS}}&=&
\delta\bm{q}_n^{\gamma(1)}|_{{\rm DIS}_{\gamma}}\cdot \bm{C}_{2,n}^{(0)}~,  \\
\frac{\alpha\alpha_s(Q^2)}{(4\pi)^2}3n_f \langle e^4 \rangle c_{2,n}^{\gamma(2)}|_{\overline{\rm MS}}&=&
\delta\bm{q}_n^{\gamma(2)}|_{{\rm DIS}_{\gamma}}\cdot
\bm{C}_{2,n}^{(0)}+\delta\bm{q}_n^{\gamma(1)}|_{{\rm DIS}_{\gamma}}\cdot \frac{\alpha_s(Q^2)}{4\pi}
\bm{C}_{2,n}^{(1)}|_{\overline{\rm MS}}~.
\label{deltaqDISgamma2}
\eea

The LO $\bm{C}_{2,n}^{(0)}$ and the NLO $\bm{C}_{2,n}^{(1)}|_{\overline{\rm MS}}$ are
written as
\bea
\bm{C}_{2,n}^{(0)}&=&\left(\langle e^2 \rangle, 0, 1 \right)^{\rm T}~,\\
\bm{C}_{2,n}^{(1)}|_{\overline{\rm MS}}&=&\left(\langle e^2 \rangle c_{2,n}^{q(1)}|_{\overline{\rm MS}},~
\langle e^2 \rangle c_{2,n}^{G(1)}|_{\overline{\rm MS}},~ c_{2,n}^{q(1)}|_{\overline{\rm MS}}\right)^{\rm T}~,
\eea
where $\langle e^2\rangle=\sum_ie_i^2/n_f$.
Now dividing the quark-charge factor $\langle e^4 \rangle$ into two parts, the flavour singlet and
nonsinglet parts, as
\begin{equation}
\langle e^4 \rangle=\langle e^2 \rangle\langle e^2 \rangle +\left(\langle e^4 \rangle-
\langle e^2 \rangle^2 \right)~,
\end{equation}
the quark sectors of the difference $\delta{\bm{q}}_n^{\gamma(1)}|_{{\rm DIS}_{\gamma}}$
at the NLO are given by
\bea
\delta q^{\gamma (1)}_{S,~ n}|_{{\rm DIS}_{\gamma}}&=&\frac{\alpha}{4\pi}3n_f\langle e^2 \rangle
c_{2,n}^{\gamma(1)}|_{\overline{\rm MS}}~, \label{deltaqS(1)}\\
\delta q^{\gamma (1)}_{NS,~ n}|_{{\rm DIS}_{\gamma}}&=&\frac{\alpha}{4\pi}3n_f\left(\langle e^4 \rangle-
\langle e^2 \rangle^2\right)
c_{2,n}^{\gamma(1)}|_{\overline{\rm MS}}~, \label{deltaqNS(1)}
\eea
while we cannot tell anything about $\delta G_n^{\gamma (1)}|_{{\rm DIS}_{\gamma}}$, since
$C_{2, n}^{G(0)}=0$~.

Actually the ${\rm DIS}_{\gamma}$ scheme was introduced from the very first so that
in this scheme the photonic coefficient function $C_2^{\gamma}$, i.e., the direct-photon
contribution to $F_2^{\gamma}$, may be absorbed into the quark
distributions but not into gluon distribution. Thus we set $\delta G_n^{\gamma}|_{{\rm DIS}_{\gamma}}=0$
in all orders. In other words, we have
\begin{equation}
G^{\gamma}(n,Q^2,P^2)|_{{\rm DIS}_{\gamma}}=G^{\gamma}(n,Q^2,P^2)|_{\overline{\rm MS}}~.
\end{equation}

At the NNLO, one obtains from Eqs.(\ref{deltaqDISgamma2})-(\ref{deltaqNS(1)})
\bea
\delta q^{\gamma (2)}_{S,~ n}|_{{\rm DIS}_{\gamma}}
&=&\frac{\alpha\alpha_s(Q^2)}{(4\pi)^2}3n_f\langle e^2 \rangle\Bigl(
c_{2,n}^{\gamma(2)}|_{\overline{\rm MS}}-c_{2,n}^{\gamma(1)}|_{\overline{\rm MS}} ~
c_{2,n}^{q(1)}|_{\overline{\rm MS}}\Bigr)~,\label{QsDISgamma}\\
\delta q^{\gamma (2)}_{NS,~ n}|_{{\rm DIS}_{\gamma}}
&=&\frac{\alpha\alpha_s(Q^2)}{(4\pi)^2}3n_f\left(\langle e^4 \rangle-\langle e^2 \rangle^2\right) \Bigl(
c_{2,n}^{\gamma(2)}|_{\overline{\rm MS}}-c_{2,n}^{\gamma(1)}|_{\overline{\rm MS}} ~
c_{2,n}^{q(1)}|_{\overline{\rm MS}}\Bigr)~.
\eea
These expressions for $\delta q^{\gamma (2)}_{S,~ n}|_{{\rm DIS}_{\gamma}}$ and $\delta q^{\gamma (2)}_{NS,~ n}|_{{\rm DIS}_{\gamma}}$
were first derived
in Ref.\cite{MVV}. (See (4.19) of Ref.\cite{MVV}).
The parameters $c_{2,n}^{q(1)}|_{\overline{\rm MS}}$ and $c_{2,n}^{\gamma(1)}|_{\overline{\rm MS}}$
are given in Ref.\cite{BB} and $c_{2,n}^{\gamma(2)}|_{\overline{\rm MS}}$ in Ref.\cite{MVV}.
They are also enumerated in Sec.III of Ref.\cite{USU}.
With the knowledge of $\delta q^{\gamma (l)}_{S,~ n}|_{{\rm DIS}_{\gamma}}$ and
$\delta q^{\gamma (l)}_{NS,~ n}|_{{\rm DIS}_{\gamma}}$ ($l=1,2$),
the parton distributions in ${\rm DIS}_{\gamma}$ scheme
$\bm{q}^\gamma(n,Q^2,P^2)|_{{\rm DIS}_{\gamma}}$ is obtained
from Eq.(\ref{SchemeDifference}) up to the NNLO.

Again there is another way to get $\bm{q}^\gamma(n,Q^2,P^2)|_{{\rm DIS}_{\gamma}}$ up to the NNLO,
once we know the expression for the moment sum rule of $F_2^\gamma(x,Q^2,P^2)$ up to the NNLO and all the quantities in the expression
are the ones calculated in $\overline {\rm MS}$ scheme.
It will be shown in Appendix \ref{MomentF_2}.

%%%%%%%%%  Behaviours of parton distributions near $x=1$ %%%%%%%%%%%%
\section{\label{LargeX}Behaviours of parton distributions near $x=1$}
\smallskip
%%%%%%%%%%%%%%%%%%%%%%%%%%%%%%%%%%%%%%%%%%%%%%%%%%%%%%%%%%%%

The behaviours of parton distributions near~$x=1$
are governed by the large-$n$ limit of those moments.
In the leading order,
parton distributions are factorisation-scheme independent. For large $n$,
the moments of the flavour singlet and nonsinglet LO quark distributions,
$q^{\gamma (0)}_{S,~ n}$ and $q^{\gamma (0)}_{NS,~ n}$, both behave as
$1/(n~{\rm ln}~n)$, while the LO gluon distribution $G_n^{\gamma (0)}$
behaves as $1/(n~{\rm ln}~n)^2$.
Thus, in $x$ space, the LO parton distributions in the virtual photon vanish for
$x \rightarrow 1$ as
\bea
 q_S^{\gamma (0)}(x,Q^2,P^2)&\approx&
\frac{\alpha}{4\pi}\frac{4\pi}{\alpha_s(Q^2)}
3 n_f\langle e^2 \rangle\frac{3}{4}~\frac{-1}{{\rm ln}~(1-x)}~, \label{QuarkLO}\\
 G^{\gamma (0)}(x,Q^2,P^2)&\approx&
\frac{\alpha}{4\pi}\frac{4\pi}{\alpha_s(Q^2)}
3n_f\langle e^2 \rangle\frac{1}{6}~\frac{-{\rm ln}~x}{{\rm ln}^2~(1-x)}~.
\eea
The behaviours of $q_{NS}^{\gamma}(x,Q^2,P^2)$ for $x \rightarrow 1$,
both in LO, NLO and NNLO are found to be always
given by the corresponding expressions for $q_{S}^{\gamma}(x,Q^2,P^2)$
with replacement of the charge factor $\langle e^2 \rangle$ with
$(\langle e^4 \rangle -\langle e^2 \rangle^2)$.

In $\overline {\rm MS}$ scheme,
the moments of the NLO parton distributions are written in large-$n$ limit as
\bea
 q_{S, n}^{\gamma(1)}\vert_{\overline {\rm MS}}
&\longrightarrow& \frac{\alpha}{4\pi}3 n_f\langle e^2 \rangle 2~\frac{{\rm
ln}~n}{n}~,      \\
G^{\gamma(1)}_n\vert_{\overline {\rm MS}}
&\longrightarrow& \frac{\alpha}{4\pi} 3 n_f\langle e^2 \rangle ~\frac{1}{n^2}~.
\eea
The leading contribution to $q_{S, n}^{\gamma(1)}\vert_{\overline {\rm MS}}$ for large $n$ comes from the (1,1) component of the term $(\bm{K}^{(1)}_n P^n_-~\frac{1}{d_-^n})$ of Eq.(\ref{qv{(1)}(t)}),
which reduces to $K^{(1),n}_S \frac{1}{d^n_-}$. For large $n$, $K^{(1),n}_S$ behaves as
$(3 n_f\langle e^2 \rangle)\frac{64}{3}\frac{{\rm ln}^2n}{n}$ while $d^n_-$ as
$\frac{32}{3}{\rm ln}~n$.
Then, in $x$ space, we have near $x=1$
\bea
   q_S^{\gamma(1)}(x,Q^2,P^2)\vert_{\overline {\rm MS}}&\approx&
\frac{\alpha}{4\pi} 3 n_f\langle e^2 \rangle 2~\Bigl[-{\rm
ln}(1-x)\Bigr]~,\label{NLOMSbar}\\
G^{\gamma(1)}(x,Q^2,P^2)\vert_{\overline {\rm MS}}&\approx&
\frac{\alpha}{4\pi} 3 n_f\langle e^2 \rangle ~\Bigl[-{\rm ln}~x\Bigr] ~.
\eea
The NLO quark distributions, both
$q_S^{\gamma(1)}(x,Q^2,P^2)\vert_{\overline {\rm MS}}$ and $q_{NS}^{\gamma(1)}(x,Q^2,P^2)\vert_{\overline {\rm MS}}$,
{\it positively} diverge as $[-{\rm ln}(1-x)]$ for $x\rightarrow 1$, while
the NLO gluon distribution, $G^{\gamma(1)}(x,Q^2,P^2)\vert_{\overline {\rm MS}}$, vanishes
as $[-{\rm ln}~x]$.

On the other hand, in ${\rm DIS}_{\gamma}$ scheme,
the moments of the NLO flavour-singlet quark distribution $q^{\gamma (1)}_{S,~ n}|_{{\rm DIS}_{\gamma}}$ is expressed in large-$n$ limit as (see Eq.(\ref{SchemeDifference})),
\bea
q^{\gamma (1)}_{S,~ n}|_{{\rm DIS}_{\gamma}}&=&
q_{S, n}^{\gamma(1)}\vert_{\overline {\rm MS}}
+\delta q^{\gamma (1)}_{S,~ n}|_{{\rm DIS}_{\gamma}}\nn\\
&\longrightarrow&\frac{\alpha}{4\pi}3 n_f\langle e^2 \rangle \Bigl[ -2~\frac{{\rm
ln}~n}{n}\Bigr]~,
\eea
since $c_{2,n}^{\gamma(1)}|_{\overline{\rm MS}}$ in Eq.(\ref{deltaqS(1)})
behaves as $(-4~{\rm ln}~n)/n$ for large $n$.
Thus we have for large $x$,
\begin{equation}
q_S^{\gamma(1)}(x,Q^2,P^2)\vert_{{\rm DIS}_{\gamma}}\approx
\frac{\alpha}{4\pi} 3 n_f\langle e^2 \rangle 2~{\rm ln}(1-x)~.
\label{NLODISgamma}
\end{equation}
The NLO distribution in ${\rm DIS}_{\gamma}$ scheme, $q_{S}^{\gamma(1)}(x,Q^2,P^2)\vert_{{\rm DIS}_{\gamma}}$, {\it negatively} diverges as $x\rightarrow 1$. This is due to the
fact that the NLO photonic coefficient function $c_2^{\gamma(1)}(x)$ (the inverse Mellin transform of $c_{2,n}^{\gamma(1)}$),
which in $\overline {\rm MS}$ becomes negative and divergent for
$x\rightarrow 1$, is absorbed into the quark distributions in ${\rm DIS}_{\gamma}$ scheme~\cite{GRV}.

The moments of the NNLO parton distributions in $\overline {\rm MS}$ scheme
behave for large $n$ as
\bea
 q_{S, n}^{\gamma(2)}\vert_{\overline {\rm MS}}
&\longrightarrow& \frac{\alpha}{4\pi}\frac{\alpha_s(Q^2)}{4\pi}3 n_f\langle e^2 \rangle
\frac{8}{9}\frac{({\rm ln}~n)^3}{n}~,      \\
G^{\gamma(2)}_n\vert_{\overline {\rm MS}}
&\longrightarrow& \frac{\alpha}{4\pi} \frac{\alpha_s(Q^2)}{4\pi} 3 n_f\langle e^2 \rangle ~\frac{17}{9}~\frac{({\rm ln}~n)^2}{n^2}~,
\eea
The leading contribution to $q_{S, n}^{\gamma(2)}\vert_{\overline {\rm MS}}$ for large $n$ comes from the (1,1) component of the term $(-\bm{K}^{(2)}_n P^n_-~\frac{1}{1-d_-^n})$ in Eq.(\ref{qv{(2)}(t)}), which reduces to $K^{(2),n}_S \frac{1}{d^n_-}$.
The large-$n$ behaviour of $K^{(2),n}_S$ is $(3 n_f\langle e^2 \rangle)\frac{256}{27}\frac{({\rm ln}n)^4}{n}$. On the other hand,
the large-$n$ behaviours of $c_{2,n}^{\gamma(2)}|_{\overline{\rm MS}}$,~ $c_{2,n}^{\gamma(1)}|_{\overline{\rm MS}}$~ and $c_{2,n}^{q(1)}|_{\overline{\rm MS}}$
are given by $-\frac{80}{9}\frac{({\rm ln}~n)^3}{n}$,
$-4\frac{{\rm ln}~n}{n} $~ and~ $\frac{8}{3}({\rm ln}~n)^2$, respectively,
and thus we see from Eq.(\ref{QsDISgamma}),
\begin{equation}
\delta q^{\gamma (2)}_{S,~ n}|_{{\rm DIS}_{\gamma}}\longrightarrow
\frac{\alpha\alpha_s}{(4\pi)^2}(3 n_f\langle e^2 \rangle)\frac{16}{9}\frac{({\rm ln}~n)^3}{n}~,
\end{equation}
for large $n$. So
we find in ${\rm DIS}_{\gamma}$ scheme,
\bea
q^{\gamma,{(2)}}_{S}\vert_{{\rm DIS}_{\gamma}}&=&
q_{S, n}^{\gamma(2)}\vert_{\overline {\rm MS}}
+\delta q^{\gamma (2)}_{S,~ n}|_{{\rm DIS}_{\gamma}}\nn\\
&\longrightarrow&
\frac{\alpha}{4\pi} \frac{\alpha_s(Q^2)}{4\pi} 3 n_f\langle e^2 \rangle
~\frac{8}{3}\frac{({\rm ln}~n)^3}{n}~.
\eea
In $x$ space, therefore, the NNLO parton distributions near $x=1$ are
\bea
   q_S^{\gamma(2)}(x,Q^2,P^2)\vert_{\overline {\rm MS}}&\approx&
\frac{\alpha}{4\pi}  \frac{\alpha_s(Q^2)}{4\pi} 3 n_f\langle e^2 \rangle \frac{8}{9}~
\Bigl[-{\rm ln}^3(1-x)\Bigr]~,\label{NNLOMSbar}\\
 q_S^{\gamma(2)}(x,Q^2,P^2)\vert_{{\rm DIS}_{\gamma}}&\approx&
\frac{\alpha}{4\pi}  \frac{\alpha_s(Q^2)}{4\pi} 3 n_f\langle e^2 \rangle \frac{8}{3}~
\Bigl[-{\rm ln}^3(1-x)\Bigr]~,\label{NNLODISgamma}\\
&&\nn \\
G^{\gamma(2)}(x,Q^2,P^2)\vert_{\overline {\rm MS}}&\approx&
\frac{\alpha}{4\pi}  \frac{\alpha_s(Q^2)}{4\pi} 3 n_f\langle e^2 \rangle \frac{17}{9}~
\Bigl[-{\rm ln} x~ {\rm ln}^2(1-x)\Bigr] ~.
\eea
It is noted that NNLO quark distributions in both $\overline{{\rm MS}}$ and ${{\rm DIS}_{\gamma}}$ schemes diverge at $x=1$ as $[-{\rm ln}^3(1-x)]$ and, furthermore, that
$q_S^{\gamma(2)}(x,Q^2,P^2)\vert_{{\rm DIS}_{\gamma}}$ rises sharper than
$q_S^{\gamma(2)}(x,Q^2,P^2)\vert_{\overline {\rm MS}}$ as $x\rightarrow 1$.
This sharp rise of $q_S^{\gamma(2)}(x,Q^2,P^2)\vert_{{\rm DIS}_{\gamma}}$ as $[-{\rm ln}^3(1-x)]$ near $x=1$ is an unexpected result. Indeed, both the NLO and NNLO photonic coefficient functions
$c_2^{\gamma(1)}(x)$ and $c_2^{\gamma(2)}(x)$ in $\overline{{\rm MS}}$ scheme
negatively diverge as $x\rightarrow 1$ (see above for the large-$n$ behaviours of
$c_{2,n}^{\gamma(1)}|_{\overline{\rm MS}}$ and $c_{2,n}^{\gamma(2)}|_{\overline{\rm MS}}$). Then, the first thought is that absorbing these photonic coefficient functions into the quark distributions
would make the ${{\rm DIS}_{\gamma}}$ quark distributions behave milder than those in
${\overline {\rm MS}}$ scheme or negatively diverge as $x\rightarrow 1$. The NLO
quark distributions in ${{\rm DIS}_{\gamma}}$ scheme work fine but not the NNLO
quark distributions. This is due to the contribution of the second term
$(-c_{2,n}^{\gamma(1)}|_{\overline{\rm MS}} ~
c_{2,n}^{q(1)}|_{\overline{\rm MS}})$ to $\delta q^{\gamma (2)}_{S,~ n}|_{{\rm DIS}_{\gamma}}$ in Eq.(\ref{QsDISgamma}).

%%%%%%%%%%%%%%%%%%%%%%%%%%%%%%
\section{\label{NumericalAnalysis}Numerical analysis}
The parton distributions in the virtual photon are recovered from their moments by the inverse Mellin transformation.
%%%%%%%%%%%%%%%%%%%%%%%%%%%%%%%%%%%
\begin{figure*}
\centering
  \begin{tabular}{c@{\hspace*{10mm}}c}
    \includegraphics[scale=0.6]{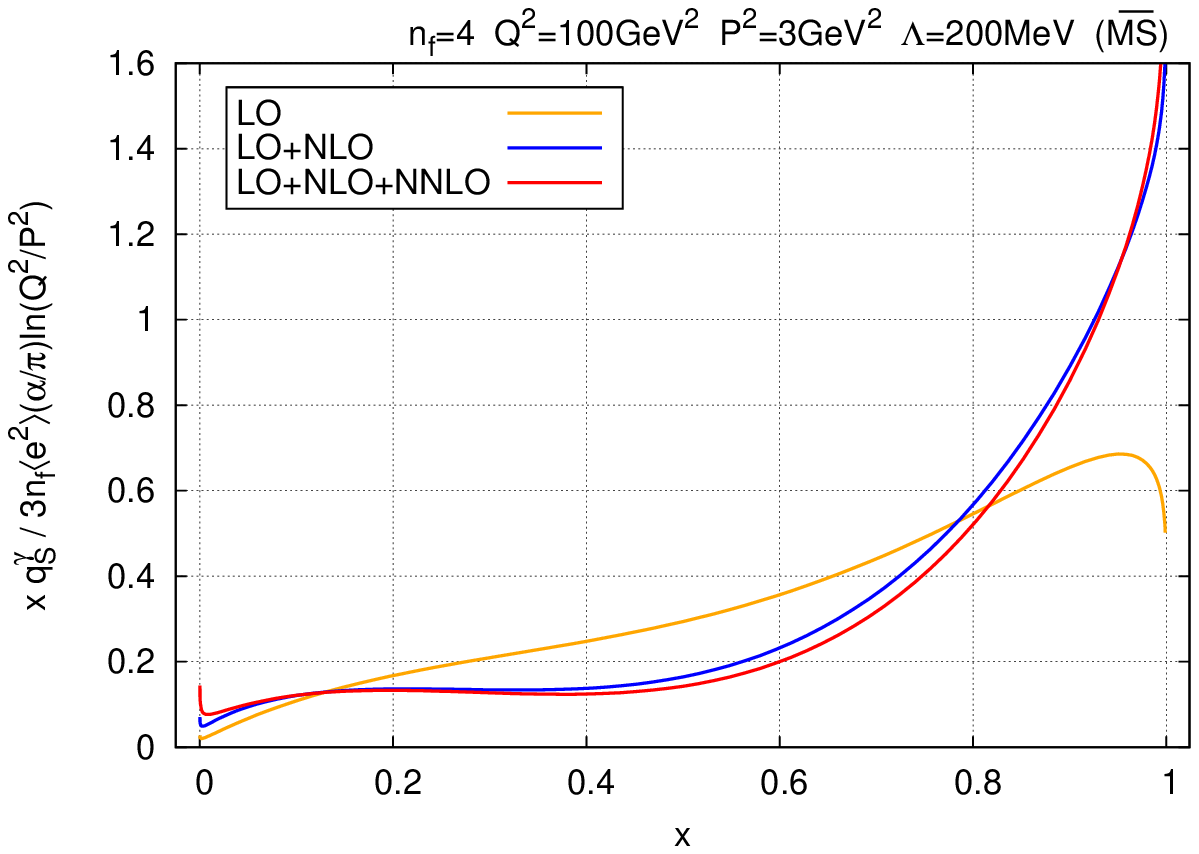}
    &
   \includegraphics[scale=0.6]{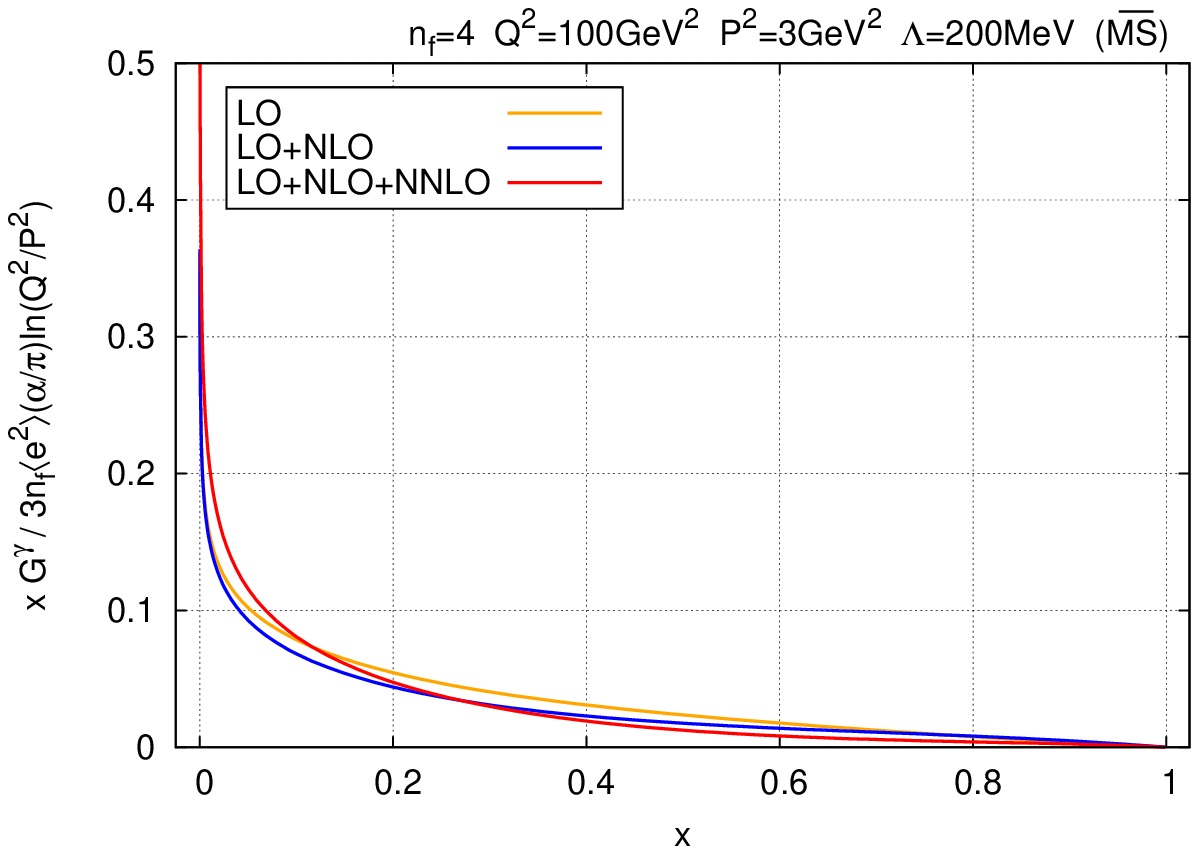}\\
(a) & (b)
  \end{tabular}
\caption{Parton distributions in the photon in $\overline{{\rm MS}}$ scheme for $n_f=4$, $Q^2=100$GeV$^2$,
$P^2=3$GeV$^2$ with $\Lambda=0.2$GeV: (a) $xq_S^\gamma(x,Q^2,P^2)|_{\overline{\rm MS}}$ and (b)
$xG^\gamma(x,Q^2,P^2)|_{\overline{\rm MS}}$.}
\label{fig1}
\end{figure*}
%%%%%%%%%%%%%%%%%%%%%%%%%%%%%%%%%%%
In Fig.\ref{fig1} we plot the parton distributions in
$\overline{{\rm MS}}$ scheme in units of $(3n_f \langle e^2\rangle \alpha/\pi){\rm ln}(Q^2/P^2)$:
(a) the singlet quark distribution $xq_S^\gamma(x,Q^2,P^2)|_{\overline{\rm MS}}$
and (b) the gluon distribution $xG^\gamma(x,Q^2,P^2)|_{\overline{\rm MS}}$. We have taken
$n_f=4$, $Q^2=100{\rm GeV}^2$, $P^2=3{\rm GeV}^2$, and the QCD scale parameter $\Lambda=0.2$GeV.
We see that both the (LO+NLO) and (LO+NLO+NNLO) curves show the similar behaviours
in almost the whole $x$ region, which means that the NNLO contribution is small.
The behaviours of these two curves, however, are quite different from the LO curve.
They lie below the LO curve for $0.2<x<0.8$, but diverge as $x\rightarrow 1$. Compared
with the quark distribution, the gluon distribution $xG^\gamma|_{\overline {\rm MS}}$ is very small in absolute value except in the small-$x$ region.
Concerning the nonsinglet quark distribution $xq_{NS}(x,Q^2,P^2)|_{\overline{\rm MS}}$, we find that when we take into
account the charge factors, such as $\langle e^2\rangle=\sum_ie_i^2/n_f$ and
$\langle e^4\rangle=\sum_ie_i^4/n_f$,
it falls on the singlet quark distribution in almost the whole $x$ region;
namely the two ``normalised" distributions
$x{\widetilde q}_S^\gamma\equiv xq_S^\gamma/\langle e^2\rangle$ and
$x{\widetilde q}_{NS}^\gamma\equiv xq_{NS}^\gamma/(\langle e^4\rangle-\langle e^2\rangle^2)$
mostly overlap except at the very small $x$ region. This situation is the same in both
$\overline {\rm MS}$ and ${\rm DIS}_{\gamma}$ schemes. The rise of the singlet quark distribution
near $x=0$ is related to the gluon distribution
which grows rapidly as $x \rightarrow 0$.

%%%%%%%%%%%%%%%%%%%%%%%%%%%%%%%%%%%
\begin{figure*}
\centering
  \begin{tabular}{c@{\hspace*{10mm}}c}
    \includegraphics[scale=0.6]{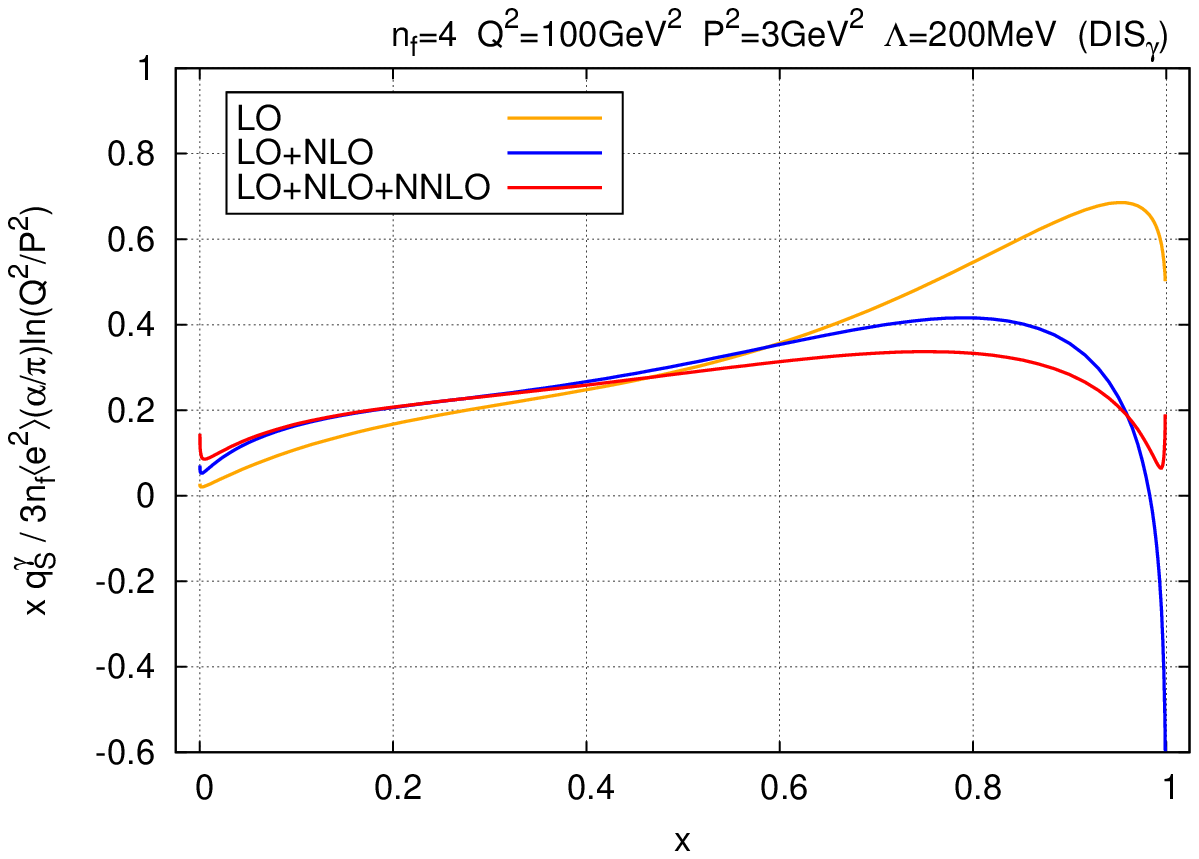}
    &
   \includegraphics[scale=0.6]{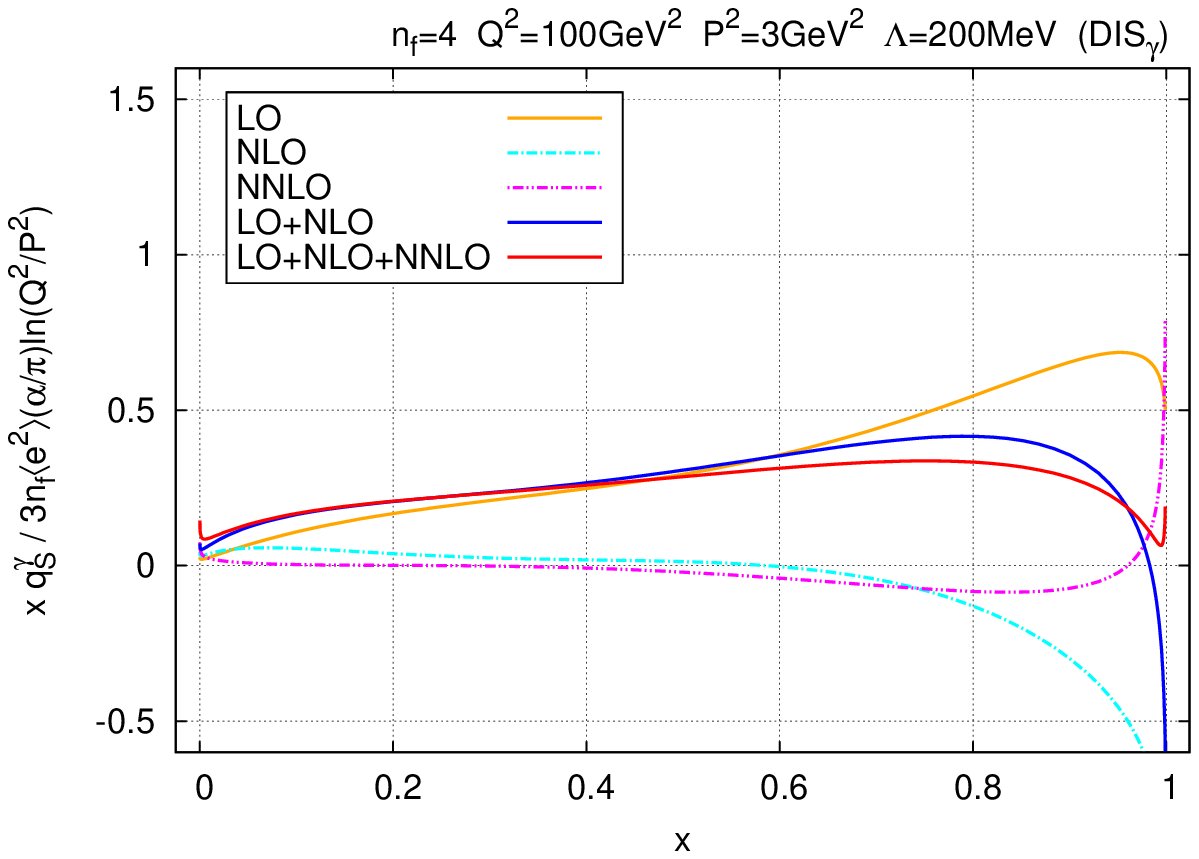}\\
(a) & (b)
  \end{tabular}
\caption{(a) Singlet quark distribution $xq_S^\gamma(x,Q^2,P^2)|_{{\rm DIS}_{\gamma}}$
in ${\rm DIS}_\gamma$ scheme for $n_f=4$, $Q^2=100$GeV$^2$,
$P^2=3$GeV$^2$ with $\Lambda=0.2$GeV. (b) LO, NLO, and NNLO contributions.
}
\label{fig2}
\end{figure*}
%%%%%%%%%%%%%%%%%%%%%%%%%%%%%%%%%%%

The parton distributions in ${\rm DIS}_{\gamma}$ scheme were analysed up to the NNLO.
In Fig.\ref{fig2}(a) we plot
 the singlet quark distribution $xq_S^\gamma(x,Q^2,P^2)|_{{\rm DIS}_{\gamma}}$
in units of $(3n_f \langle e^2\rangle \alpha/\pi){\rm ln}(Q^2/P^2)$.
Again we chose $n_f=4$, $Q^2=100{\rm GeV}^2$, $P^2=3{\rm GeV}^2$, and $\Lambda=0.2$GeV.
The three curves (LO, LO+NLO, LO+NLO+NNLO) rather overlap below $x=0.6$.
Absorbing the photonic coefficient function $C_2^\gamma$ into the quark distributions
in ${\rm DIS}_{\gamma}$ scheme has an effect on their large-$x$ behaviours:
Unlike the $\overline {\rm MS}$ scheme, the (LO+NLO) curve goes under the LO curve at $x\approx 0.6$ and the difference
between the two grows as $x\rightarrow 1$. Adding the NNLO contribution makes
the difference bigger at large $x$ except near $x=1$. At very close to $x=1$
the (LO+NLO+NNLO) curve shows a sudden surge. In order to see the details,
we plot in Fig.{\ref{fig2}}(b), the NLO and NNLO contributions in ${\rm DIS}_{\gamma}$ scheme. We observe that the NLO contribution is large and negative for $x>0.8$, while
the NNLO contribution remains to be very small until very close to $x=1$ and then blows up.
The behaviours near $x=1$ of $q_S^{\gamma(1)}(x,Q^2,P^2)\vert_{{\rm DIS}_{\gamma}}$
and $q_S^{\gamma(2)}(x,Q^2,P^2)\vert_{{\rm DIS}_{\gamma}}$ were discussed in
Sec.\ref{LargeX}.

Finally, the gluon distribution $xG^\gamma(x,Q^2,P^2)|_{{\rm DIS}_{\gamma}}$ is the same as
$xG^\gamma(x,Q^2,P^2)|_{\overline {\rm MS}}$.

%%%%%%%%%%%%%%%%%%%%%%%%%%%%%%%

%%%%%%%%%%%%%%%%%%%%%%%

\section{Conclusions}

We have analysed the parton distributions in the virtual photon target
which are predicted entirely up to the NNLO in perturbative QCD.
Parton distributions are dependent on the scheme which is employed to factorise
structure functions into parton distributions and
coefficient functions. The virtual photon target serves as a good testing ground
for examining the behaviours of the parton distributions and their
factorisation-scheme dependences.
We have studied the quark and gluon distributions in two factorisation schemes,
namely, $\overline {\rm MS}$ and ${\rm DIS}_{\gamma}$ schemes.

We see from Figs.\ref{fig1}(a) and \ref{fig2}(a) that
(LO+NLO) and (LO+NLO+NNLO) curves for
the quark distribution $xq_S^\gamma(x,Q^2,P^2)$ show quite different behaviours in two schemes, especially in the large-$x$ region. From the viewpoint of ``perturbative stability",
the ${\rm DIS}_{\gamma}$ scheme
gives a more appropriate behaviour for $xq_S^\gamma(x,Q^2,P^2)$ than $\overline {\rm MS}$.
The gluon distribution $xG^\gamma(x,Q^2,P^2)$ is the same in both schemes and is predicted to be very
small in absolute value except in the small-$x$ region.
Finally, we observe that the (LO+NLO+NNLO) curve for $xq_S^\gamma(x,Q^2,P^2)|_{{\rm DIS}_{\gamma}}$ shows a sudden surge at very close to $x=1$. Although the NLO contribution,
${q_S^{\gamma (1)}}|_{{\rm DIS}_{\gamma}}$, negatively diverges as ${\rm ln} (1-x)$ for
$x\rightarrow 1$, the NNLO ${q_S^{\gamma (2)}}|_{{\rm DIS}_{\gamma}}$ diverges
positively as $[-{\rm ln}^3 (1-x)]$.
This may hint a necessity of considering the resummation
for parton distributions and also for the photon structure function $F_2^\gamma(x,Q^2,P^2)$ itself
near $x=1$~\cite{Catani}.

%%%%%%%%%%%%%%%%%%%%%%%%%%

\begin{acknowledgments}
We thank Stefano Catani for valuable discussions. This work is supported in part by Grant-in-Aid for
Scientific Research (C) from the Japan Society for the Promotion
of Science No.18540267.
\end{acknowledgments}

\newpage
\appendix

\section{\label{Parameters}Parameters in $\bm{q}^{\gamma(0)}(t)$, $\bm{q}^{\gamma(1)}(t)$
and $\bm{q}^{\gamma(2)}(t)$}
We give here some important information on the parameters which appear in
$\bm{q}^{\gamma(0)}(t)$, $\bm{q}^{\gamma(1)}(t)$
and $\bm{q}^{\gamma(2)}(t)$ in Eqs.(\ref{qv{(0)}(t)})-(\ref{qv{(2)}(t)}).
They are all calculated in $\overline{\rm MS}$ scheme.
We introduce the following quark charge factors:
\bea
&&\langle e^2 \rangle=\frac{1}{n_f}\sum^{n_f}_{i=1}e^2_i~, \qquad
\langle e^4 \rangle=\frac{1}{n_f}\sum^{n_f}_{i=1}e^4_i~,\nn\\
&&f_S=3n_f\langle e^2 \rangle~,\qquad f_{NS}=3n_f\Bigl(\langle e^4 \rangle
-\langle e^2 \rangle^2\Bigr)~.
\eea

\subsection{\label{InitialConditions}Initial conditions for parton distributions:
${\widetilde{\bm{A}}}_n^{(1)}$ and ${\widetilde{\bm{A}}}_n^{(2)}$}

The elements of the row vector ${\widetilde{\bm{A}}}_n^{(l)}$ are expressed as
${\widetilde{\bm{A}}}_n^{(l)}=\Bigl({\widetilde A}_{n}^{S(l)}, {\widetilde A}_{n}^{G(l)}, {\widetilde
A}_{n}^{NS(l)} \Bigr)$, with $l=1,2$. Then we have at one-loop level,
\begin{equation}
{\widetilde A}_n^{S(1)}=f_S H_q^{(1)}(n)~, \quad
{\widetilde A}_n^{G(1)}=0~, \quad
{\widetilde A}_n^{NS(1)}=f_{NS} H_q^{(1)}(n)~,
\label{PhotonHadronicMatOneLoop}
\end{equation}
where the explicit expression of $H_q^{(1)}(n)$ is given in Eq.(3.30) of Ref.\cite{USU}. The
elements at two-loop level are
\begin{equation}
{\widetilde A}_n^{S(2)}=f_S H_q^{(2)}(n)~,\quad
{\widetilde A}_n^{G(2)}=f_S H_G^{(2)}(n)~,\quad
{\widetilde A}_n^{NS(2)}=f_{NS} H_q^{(2)}(n)~,
\label{PhotonQuarkMatTwoLoop}
\end{equation}
where $H_q^{(2)}(n)$ and $H_G^{(2)}(n)$ are given in Eqs.(3.39) and (3.42) of Ref.\cite{USU},
respectively.

\subsection{Anomalous dimensions}
The one-, two- and three-loop anomalous dimensions for the hadronic sector,
${\widehat \gamma}_{n}^{(0)}$,
${\widehat \gamma}_{n}^{(1)}$ and ${\widehat \gamma}_{n}^{(2)}$~\cite{MVV1,MVV2}, respectively, are already in literature. The eigenvalues $\lambda_i^n$ of the one-loop
anomalous dimension matrix ${\widehat \gamma}_{n}^{(0)}$ and
the corresponding projection operators $P_i^n$ which appeared in Eqs.
(\ref{EigenProjector})-(\ref{Projector}) are given in Ref.\cite{BB}.
Also the one- and two-loop photonic anomalous dimensions, $\bm{K}_n^{(0)}$ and $\bm{K}_n^{(1)}$, are already known~\cite{BB, FP2, GRV}. But
concerning
the three-loop anomalous dimensions $K_S^{{(2),n}}$, $ K_G^{{(2),n}}$ and $K_{NS}^{{(2),n}}$,
the exact expressions have not been in literature yet, but approximate ones were given~\cite{MVV3}.
It is remarked there that the precision is within about 0.1\% or less.
The approximate expressions for $K_S^{{(2),n}}$, $K_{NS}^{{(2),n}}$ and $ K_G^{{(2),n}}$
are
\bea
K_S^{{(2),n}}&\approx&K_{S\ \rm approx}^{{(2),n}} \equiv-f_S~ 2\Bigl\{E_{{\rm
ns}\gamma}^{\rm approx}(n) + E_{{\rm ps}\gamma}(n)\Bigr\}~, \nn\\
 K_G^{{(2),n}}&\approx&K_{G\ \rm approx}^{{(2),n}} \equiv-f_S~ 2E_{G\gamma}^{\rm
approx}(n)~, \label{EgApprox}\\
K_{NS}^{{(2),n}}&\approx&K_{NS\ \rm approx}^{{(2),n}} \equiv -f_{NS}~
2E_{{\rm ns}\gamma}^{\rm approx}(n)~,\nn
\eea
where the explicit expressions of $E_{{\rm ns}\gamma}^{\rm approx}(n)$, $ E_{{\rm ps}\gamma}(n)$,
and $E_{G\gamma}^{\rm approx}(n)$ are given, respectively, in Eqs.(B2), (B4) and (B3) of Ref.\cite{USU}.

%%%%%%%%%%%%%%%%%%%%%%%%%%%%%%%%%
\section{\label{MomentF_2}Another way to find parton distributions in ${\overline{\rm MS}}$ and ${\rm DIS}_{\gamma}$ scheme}

%%%%%%%%%%%%%%%%%%%%%%
Once we know the expression for the moment sum rule of $F_2^\gamma(x,Q^2,P^2)$
obtained up to the NNLO corrections \cite{USU} and all the quantities in the expression
are the ones calculated in $\overline {\rm MS}$ scheme, then there is an easy way to find the parton
distributions in the virtual photon in both ${\overline{\rm MS}}$ and ${{\rm
DIS}_{\gamma}}$ schemes up to the NNLO. In the following the equation numbers
correspond to those in Ref.\cite{USU}.
\begin{description}
\item{(i)} Parton distributions up to the NNLO in ${\overline{\rm MS}}$ scheme\\
In the moment sum rule of $F_2^\gamma(x,Q^2,P^2)$ given by Eqs.(2.29)-(2.37),
we set ${\bm{C}}_{2,n}^{(1)}={\bm{C}}_{2,n}^{(2)}=\bm{0}$ and $C_{2,n}^{\gamma (1)}=
C_{2,n}^{\gamma (2)}=0$. In addition, we put $\bm{C}_{2,n}^{(0)}=(1,0,0)^{\rm T}$,
then we obtain $q^{\gamma}_{S}(n,Q^2,P^2)|_{\overline{\rm MS}}$. Similarly, when we put
$\bm{C}_{2,n}^{(0)}=(0,1,0)^{\rm T}$ and $\bm{C}_{2,n}^{(0)}=(0,0,1)^{\rm T}$,
we obtain $G^{\gamma}(n,Q^2,P^2)|_{\overline{\rm MS}}$ and
$q^{\gamma}_{NS}(n,Q^2,P^2)|_{\overline{\rm MS}}$~, respectively.

\item{(ii)} Parton distributions up to the NNLO in ${\rm DIS}_{\gamma}$ scheme\\
For the gluon distribution, we have
$G^{\gamma}(n,Q^2,P^2)|_{{\rm DIS}_{\gamma}}=G^{\gamma}(n,Q^2,P^2)|_{\overline{\rm MS}}$~.

\end{description}
\bi
\item $q^{\gamma}_{NS}(n,Q^2,P^2)|_{{\rm DIS}_{\gamma}}$ is obtained as follows:\\
 Writing~ $q^{\gamma}_{NS}|_{{\rm DIS}_{\gamma}}$ ~as~
$q^{\gamma}_{NS}|_{{\rm DIS}_{\gamma}}=q^{\gamma,{(0)}}_{NS}|_{{\rm
DIS}_{\gamma}}+q^{\gamma,{(1)}}_{NS}|_{{\rm DIS}_{\gamma}}+q^{\gamma,{(2)}}_{NS}|_{{\rm
DIS}_{\gamma}}$,
 \bi

\item $q^{\gamma,{(0)}}_{NS}|_{{\rm DIS}_{\gamma}}/\Bigl[\frac{\alpha}{8\pi\beta_0} \Bigr]
\Bigl[\frac{4\pi}{\alpha_s(Q^2)}\Bigr]$ is obtained from the term
${\cal{L}}_{NS}^n$ in Eq.(2.30), which is the same with $q^{\gamma,{(0)}}_{NS}|_{\overline {MS}}$.

\item $q^{\gamma,{(1)}}_{NS}|_{{\rm DIS}_{\gamma}}/\Bigl[\frac{\alpha}{8\pi\beta_0} \Bigr]$ is
obtained from the sum of the terms
${\cal{A}}_{NS}^n$, ${\cal{B}}_{NS}^n$ and ${\cal{C}}^n$ in Eqs.(2.31)-(2.33), where we put, ${\bm C}_{2,n}^{(0)}=(0,0,1)^T$, ${\bm C}_{2,n}^{(1)}=\bm 0$
and $C_{2,n}^{\gamma(1)}$ is replaced with
$C_{2,n}^{\gamma(1)}\times
\frac{\langle e^4 \rangle- \langle e^2 \rangle^2}{\langle e^4 \rangle}$.

\item $q^{\gamma,{(2)}}_{NS}|_{{\rm DIS}_{\gamma}}/\Bigl[\frac{\alpha}{8\pi\beta_0} \Bigr]
\Bigl[\frac{\alpha_s(Q^2)}{4\pi}\Bigr]$ is obtained
from the sum of the terms
${\cal{D}}_{NS}^n$, ${\cal{E}}_{NS}^n$, ${\cal{F}}_{NS}^n$ and ${\cal{G}}^n$
in Eqs.(2.34)-(2.37), where we
put, ${\bm C}_{2,n}^{(0)}=(0,0,1)^T$, ${\bm C}_{2,n}^{(1)}=\bm 0$,
${\bm C}_{2,n}^{(2)}=\bm 0$ and $C_{2,n}^{\gamma(2)}$ is replaced with \quad
$\Bigl(C_{2,n}^{\gamma(2)}- C_{2,n}^{\gamma(1)} C_{2,n}^{NS(1)} \Bigr)\times
\frac{\langle e^4 \rangle- \langle e^2 \rangle^2}{\langle e^4 \rangle}$.
\ei

\item $q^{\gamma}_{S}(n,Q^2,P^2)|_{{\rm DIS}_{\gamma}}$ is obtained as follows:\\
 Writing~ $q^{\gamma}_{S}|_{{\rm DIS}_{\gamma}}$ ~as~
$q^{\gamma}_{S}|_{{\rm DIS}_{\gamma}}=q^{\gamma,{(0)}}_{S}|_{{\rm
DIS}_{\gamma}}+q^{\gamma,{(1)}}_{S}|_{{\rm DIS}_{\gamma}}+q^{\gamma,{(2)}}_{S}|_{{\rm
DIS}_{\gamma}}$,
 \bi

\item $q^{\gamma,{(0)}}_{S}|_{{\rm DIS}_{\gamma}}/\Bigl[\frac{\alpha}{8\pi\beta_0} \Bigr]
\Bigl[\frac{4\pi}{\alpha_s(Q^2)}\Bigr]$ is obtained from the terms
$\Bigl({\cal{L}}_{+}^n+{\cal{L}}_{-}^n\Bigl)/\langle e^2 \rangle$
in Eq.(2.30), which is the same as
$q^{\gamma,{(0)}}_{S}|_{\overline {MS}}$. Or we can obtain from the sum of the terms
$\sum_i{\cal{L}}_i^n$, where we put, ${\bm C}_{2,n}^{(0)}=(1,0,0)^T$.

\item $q^{\gamma,{(1)}}_{S}|_{{\rm DIS}_{\gamma}}/\Bigl[\frac{\alpha}{8\pi\beta_0} \Bigr]$ is
obtained from the sum of the terms
$\sum_i{\cal{A}}_i^n$, $\sum_i{\cal{B}}_i^n$ and ${\cal{C}}^n$
in Eqs.(2.31)-(2.33), where we
put, ${\bm C}_{2,n}^{(0)}=(1,0,0)^T$, ${\bm C}_{2,n}^{(1)}=\bm 0$
and $C_{2,n}^{\gamma(1)}$ is replaced with
$C_{2,n}^{\gamma(1)}\times
\frac{\langle e^2 \rangle}{\langle e^4 \rangle}$.

\item $q^{\gamma,{(2)}}_{S}|_{{\rm DIS}_{\gamma}}/\Bigl[\frac{\alpha}{8\pi\beta_0} \Bigr]
\Bigl[\frac{\alpha_s(Q^2)}{4\pi}\Bigr]$ is obtained
from the sum of the terms
$\sum_i{\cal{D}}_i^n$, $\sum_i{\cal{E}}_i^n$, $\sum_i{\cal{F}}_i^n$ and ${\cal{G}}^n$
in Eqs.(2.34)-(2.37), where we
put, ${\bm C}_{2,n}^{(0)}=(1,0,0)^T$, ${\bm C}_{2,n}^{(1)}=\bm 0$,
${\bm C}_{2,n}^{(2)}=\bm 0$ and $C_{2,n}^{\gamma(2)}$ is replaced with \quad
$\Bigl(C_{2,n}^{\gamma(2)}- C_{2,n}^{\gamma(1)} C_{2,n}^{NS(1)} \Bigr)\times
\frac{\langle e^2 \rangle}{\langle e^4 \rangle}$.\qquad Note $C_{2,n}^{S(1)}/
\langle e^2 \rangle =C_{2,n}^{NS(1)}$.
\ei

\ei
\bigskip

%\newpage %Just because of unusual number of tables stacked at end

\newpage
 %\baselineskip 17pt

%%%%%%%%%%%%%%%%%%%%%%%%%%%%%%%%

\end{document}